\documentclass[showpacs,amssymb,twocolumn,floatfix,pre,aps]{revtex4-1}

\usepackage{amssymb,amsfonts,bm}
\usepackage{graphics,graphicx}

\begin{document}

\title{Wigner-Crystal Formulation of Strong-Coupling Theory 
for Counter-ions Near Planar Charged Interfaces}

\author{Ladislav \v{S}amaj}
\altaffiliation[On leave from ]
{Institute of Physics, Slovak Academy of Sciences, Bratislava, Slovakia}
\author{Emmanuel Trizac}
\affiliation{Laboratoire de Physique Th\'eorique et Mod\`eles Statistiques, 
UMR CNRS 8626, Universit\'e Paris-Sud, 91405 Orsay, France}

\begin{abstract}
We present a new analytical approach to the strong electrostatic coupling regime (SC),
that can be achieved equivalently at low temperatures, high charges,
low dielectric permittivity etc. Two geometries are analyzed in detail: 
one charged wall first, and then, two parallel walls  at small distances, 
that can be likely or oppositely charged. In all cases, one type of mobile counter-ions
only is present, and ensures electroneutrality (salt free case). 
The method is based on a systematic expansion around the ground state 
formed by the two-dimensional Wigner crystal(s) of counter-ions at 
the plate(s). 
The leading SC order stems from a single-particle theory, and coincides
with the virial SC approach that has been much studied in the last 10 years.
The first correction has the functional form of the virial SC prediction, 
but the prefactor is different.
The present theory is free of divergences and the obtained results,
both for symmetrically and asymmetrically charged plates, are in excellent 
agreement with available data of Monte-Carlo simulations under strong and 
intermediate Coulombic couplings. 
All results obtained  represent relevant improvements over 
the virial SC estimates.
The present SC theory starting from the Wigner crystal and therefore coined
Wigner SC, sheds light on anomalous 
phenomena like the counter-ion mediated like-charge attraction, and 
the opposite-charge repulsion.
\end{abstract}


\date{\today} 

\maketitle

\section{Introduction}
Understanding effective equilibrium interactions between two charged mesoscopic bodies 
immersed in a solution, is essential in 
various fields of colloid science, from physics \cite{Ben95} to biochemistry 
\cite{Jonsson99}. 
References \cite{Gelbart00,HL00,Belloni00,Levin02,Grosberg02,Messina09,NKNP10} 
offer a general overview.
A breakthrough in the field was achieved when it was realized in the 1980s,
from numerical evidences,
that equivalently
charged surfaces may effectively attract each other, under strong enough 
Coulombic couplings. Such couplings can be realized in practice by increasing 
the valency of the counter-ions involved \cite{Guldbrand84}. This ``anomalous''
like-charge attraction explains 
the formation of DNA condensates \cite{Bloomfield96} or 
aggregates of colloidal particles \cite{Linse99}.
A complementary interesting although simpler to
rationalize problem is the possibility of an effective 
repulsion between two plates with opposite uniform surface charges.

The weak-coupling limit is described by the Poisson-Boltzmann (PB) 
mean-field approach.
Formulating the Coulomb problem as a field theory, the PB equation
can be viewed as the first-order term of a systematic expansion in loops
\cite{Attard88}.
While the like-charge attraction is not predicted by the PB theory
\cite{Neu99,Sader,T00,Andelman06}, the opposite-charge repulsion can occur already
in the mean-field treatment \cite{Parsegian72,Paillusson11},
since it is merely an entropic effect with a large cost for confining
particles in a small volume.

A remarkable theoretical progress has been made during the past decade in the opposite
strong-coupling (SC) limit, formulated initially for a single wall or two parallel 
walls at small separation. 
The topic was pioneered by Rouzina and Bloomfield \cite{Rouzina96} and
developed further by Shklovskii, Levin with collaborators \cite{Shklovskii,Levin02}. 
An essential aspect is that counter-ions form two-dimensional (2D) 
highly correlated layers at charged walls at temperature $T=0$. For small
but non vanishing temperatures, the structure of interfacial counter-ions 
remains close to its ground-state counterpart.

Within the field-theoretical formulation, which has been put forward by Netz and 
collaborators in \cite{Moreira00,Netz01}, the leading SC behavior is 
a single-particle theory in the potential of the charged wall(s).
Next correction orders as obtained as a virial or fugacity expansion
in inverse powers of the coupling constant $\Xi$, defined below;
we refer to this approach as the virial strong-coupling (VSC) theory.
The method requires a renormalization of infrared divergences
via the electroneutrality condition.
A comparison with Monte Carlo (MC) simulations 
\cite{Moreira00} indicated
the adequacy of the VSC approach to capture the leading large-$\Xi$
behaviour of the density profile, which was an important achievement in the field.
The first correction has the right functional form in space but
an incorrect prefactor, whose values even depart further from the MC ones
as the coupling constant $\Xi$ grows.
This deficiency was attributed by the authors to the existence of 
an infinite sequence of higher-order logarithmic terms in the fugacity 
which have to be resummed to recover the correct value of the prefactor.
The {\em leading} order of the VSC theory was generalized to non-symmetrically 
charges plates \cite{Kanduc08,Paillusson11}, image charge effects \cite{Kanduc07}, 
presence of salt \cite{Kanduc10} and to various curved
(spherical and cylindrical) geometries, for a review, see \cite{Naji05}.
Beyond Refs. \cite{Moreira00},
several investigations assessed numerically the adequacy of the
leading order VSC approach
\cite{Moreira00,Najicyl05,Kanduc07,Kanduc08,Dean09,Kanduc10}.

Since the coupling constant $\Xi\propto 1/T^2$, the zero temperature is 
contained in the VSC approach as the limit $\Xi\to\infty$. This question
requires some care though, since a natural rescaled distance $\widetilde{z}=z/\mu$ in the direction
perpendicular to the plate(s) is set by the Gouy-Chapman length 
$\mu\propto T$, which tends to zero as $T\to 0$. 
From this point of view, the VSC method can be seen as a low-temperature theory
approaching $T=0$ under a special spatial scaling of particle coordinates.
One of the restrictions is the applicability of the theory to small 
(rescaled) distances between the charged plates.
There exist other possibilities to approach the zero temperature limit.
One of them is to construct an expansion in $\Xi$ around the limit 
$\Xi\to\infty$, under the fixed ratio of the distance $d$ (in the two plate problem) and 
the lattice spacing $a$ of the Wigner crystal formed at $T=0$.
The low-temperature theory proposed by Lau et al. \cite{Lau00},
can be considered in some respect as being of this kind.
The considered model consists of two staggered hexagonal Wigner crystals 
of counter-ions condensed on the plates; the particles are not allowed 
to move in the slab between the plates.
The attraction between the plates at zero and non-zero temperatures,
which results from the interaction of the staggered Wigner crystals 
and from the particle fluctuations, can be computed.
Since the particles are not allowed to leave their Wigner plane, the
counter-ion profile between the two plates is trivial and there is no need
for a spatial scaling.
Such a model is interesting on its own, but has a restricted applicability
to realistic systems of counterions because the particles are assumed to 
stick to the plates. 
This assumption may be perhaps acceptable at large distances between plates, 
but discards from the outset the excitations that are relevant at small 
distances, where the counter-ions unbind from the interfaces (see e.g. 
\cite{Netz01,Moreira00} and the analysis below). 

An interpolation between the Poisson-Boltzmann (low $\Xi$) and SC regimes (high $\Xi$), 
based on the idea of a ``correlation hole'',  was the subject of 
a series of works \cite{Nordholm84,Chen06,Hatlo10}.
The correlation hole was specified empirically in Refs. \cite{Chen06} 
and self-consistently, as an optimization condition for 
the grand partition function, in \cite{Hatlo10}.
An interesting observation in \cite{Hatlo10}, corroborated by a comparison
with the MC simulations, was that the first correction
in the SC expansion is proportional to $1/\sqrt{\Xi}$, 
and not to $1/\Xi$ as suggested by the VSC theory. 
Our exact expansion below shows that indeed, the first correction
scales like $1/\sqrt{\Xi}$.

Recently, for the geometries of one plate and two equivalently 
charged plates with counter-ions only, we proposed another type of 
SC theory \cite{Samaj11a}. 
It is based on a low-temperature expansion in particle deviations
around the ground state formed by the 2D Wigner crystal of counter-ions at 
the plate(s).
The approach points to the primary importance of the structure of the ground
state, a point emphasized by some authors, see e.g. \cite{Levin99}. Our starting
point therefore resembles that of Ref. \cite{Lau00},
but in the subsequent analysis, 
the particles vibrations around their Wigner lattice positions are allowed
along all directions, including the direction perpendicular to 
the crystal plane along which the particle density varies in a nontrivial way.
The theory is formulated in the set-up of the original VSC approach:
An SC expansion around the same limit $\Xi\to\infty$ is made, together 
with the same scaling of the coordinate in the direction perpendicular 
to the plate(s), $\widetilde{z}=z/\mu$.
Since the formation of the Wigner crystal is the basic ingredient from 
which the method starts, we shall refer to it as the WSC theory.
Its leading order stems from a single-particle theory, and is identical to the 
leading order obtained in the VSC approach. In the present planar geometry, 
both WSC and VSC differ beyond the leading order, when the first correction 
is considered. In this respect, in assessing the physical relevance
of WSC and VSC, comparison to ``exact'' numerical data is essential.
Remarkably, the first WSC correction has the functional form in space of 
the VSC prediction, but the prefactor is different: 
Its $1/\sqrt{\Xi}$ dependence on the coupling parameter and 
the value of the corresponding prefactor are in excellent agreement with 
available data of MC simulations, while the VSC prediction is off by several
orders of magnitude under strong Coulombic couplings 
\cite{Moreira00}.  
Unlike the VSC theory, the WSC expansion is free of divergences, without
any need for a renormalization of parameters. 
The WSC expansion turns out to be in inverse powers of $\sqrt{\Xi}$, and not
of $\Xi$ like in the case of the VSC expansion.  
Due to its relatively simple derivation and algebraic structure, 
the WSC method has a potential applicability to a large variety of 
SC phenomena. In particular, the WSC can be worked out beyond the
leading order for asymmetric plates, which, to our knowledge,
was not done at the VSC level, possibly due to the technical
difficulty to overcome. 
The specific 2D Coulomb systems with logarithmic pair interactions 
were treated at WSC level in Ref. \cite{Samaj11b}.

In this paper, we aim at laying solid grounds for 
the WSC method.
We develop the mathematical formalism initiated in Ref. \cite{Samaj11a}, 
which is based on a cumulant expansion, to capture systematically 
vibrations of counter-ions around their Wigner crystal positions. 
This formalism enables us to deal, in the leading order plus the first
correction, also with asymmetric, likely or oppositely charged plates.
The results obtained are in remarkable agreement with MC data, for large 
as well as intermediate values of the coupling parameter $\Xi$.   

The paper is organized as follows.
The one-plate geometry is studied in Sec. \ref{sec:oneplate}.
An analysis is made of counter-ions vibrations around their ground-state 
positions in the Wigner crystal, along both transversal and 
longitudinal directions with respect to the plate surface.
The cumulant technique, providing us with the WSC expansions of the particle 
density profile in powers of $1/\sqrt{\Xi}$, is explained in detail.
Sec. \ref{sec:twoplates} deals with the geometry of two parallel plates at small separation.
The cumulant technique is first implemented for equivalently charged plates
and afterwards for asymmetrically charged plates.
In the case of the opposite-charged plates, the WSC results for the pressure 
are in agreement with MC simulations for small plate separations and 
lead to the correct (nonzero) large-distance asymptotics. 
In the case of the like-charged plates, the accurate WSC results for
the pressure are limited to small plate separations.
All obtained results represent an essential improvement over the VSC estimates.
Concluding remarks are given in Sec. \ref{sec:concl}.

Before we embark on our study, a semantic point is in order.
Some authors refer to the VSC approach as the ``SC theory''. 
Clearly, the VSC route is not the only theory that can
be put forward to describe the strong coupling regime. 
In what follows, the SC limit refers to $\Xi\to \infty$,
and we carefully discriminate between VSC and WSC predictions,
that will both be tested against Monte Carlo data.

\section{One-plate geometry}
\label{sec:oneplate}

\subsection{Definitions and notations}

\begin{figure}[htb]
\begin{center}
\includegraphics[width=0.45\textwidth]{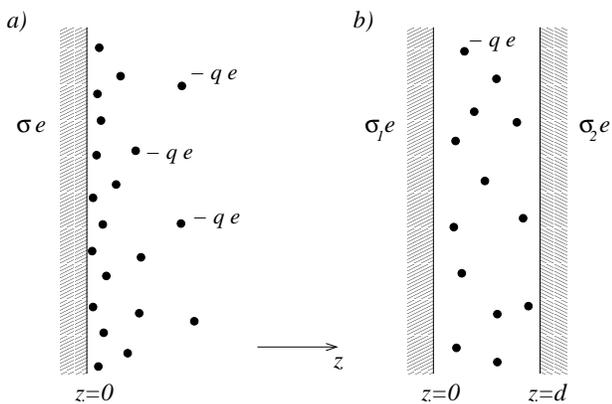}
\caption{The two geometries considered: a) one plate; b) two parallel plates
at distance $d$. The neutralizing
counter-ions have charge $-q e$.}
\label{fig:geometry} 
\end{center}
\end{figure}

We start with the one plate problem in the 3D Euclidean space
of points ${\bf r}=(x,y,z)$ pictured in Fig. \ref{fig:geometry}a.
In the half-space $\Lambda'=\{ {\bf r},z<0\}$, there is a hard wall 
of dielectric constant $\varepsilon$ which is impenetrable to particles.
A uniform surface-charge density $\sigma e$, $e$ being the elementary charge 
and $\sigma>0$, is fixed at the wall surface $\Sigma$ localized at $z=0$. 
The $q$-valent counter-ions (classical point-like particles) of 
charge $-q e$, immersed in a solution of dielectric constant $\varepsilon$, 
are confined to the complementary half-space $\Lambda=\{ {\bf r},z\ge 0\}$. 
In this work, we consider the homogeneous dielectric case only, 
without electrostatic image forces.
The system is in thermal equilibrium at the inverse temperature
$\beta = 1/(k_{\rm B}T)$.

The potential energy of an isolated counter-ion at distance $z$ from
the wall is, up to an irrelevant constant, given by
\begin{equation} \label{eq:potentialenergy}
E(z) = \frac{2\pi q e^2 \sigma}{\varepsilon} z .
\end{equation}
The system as a whole is electroneutral; denoting the (infinite) number
of counter-ions by $N$ and the (infinite) area of the wall surface by 
$\vert\Sigma\vert$, the electroneutrality condition reads
\begin{equation} \label{eq:electroneutrality}
q N = \sigma \vert\Sigma\vert .
\end{equation} 

There are two relevant length scales describing, in Gaussian units,  
the interaction of counter-ions with each other and with the charged surface.
The Bjerrum length
\begin{equation}
\ell_{\rm B} = \frac{\beta e^2}{\varepsilon}
\end{equation}  
is the distance at which two unit charges interact with thermal energy 
$k_{\rm B}T$.
The Gouy-Chapman length
\begin{equation}
\mu = \frac{1}{2\pi q \ell_{\rm B}\sigma}
\end{equation}
is the distance from the charged wall at which an isolated counter-ion
has potential energy (\ref{eq:potentialenergy}) equal to thermal energy
$k_{\rm B}T$.
The $z$ coordinate of particles will be usually expressed in units of $\mu$,
\begin{equation}
\widetilde{z} = \frac{z}{\mu} .
\end{equation}
The dimensionless coupling parameter $\Xi$, quantifying the strength of 
electrostatic correlations, is defined as the ratio
\begin{equation}
\Xi = \frac{q^2 \ell_{\rm B}}{\mu} = 2\pi q^3 \ell_{\rm B}^2 \sigma .
\end{equation}
The strong-coupling regime $\Xi\gg 1$ corresponds to either low temperatures, 
or large valency $q$ or surface charge $\sigma e$. 

The counter-ion averaged density profile $\rho(z)$ depends on the distance
$z$ from the wall.
It will be considered in the rescaled form
\begin{equation}
\widetilde{\rho}(\widetilde{z}) \equiv \frac{\rho(\mu\widetilde{z})}{
2\pi\ell_{\rm B}\sigma^2} .
\end{equation}
The electroneutrality condition (\ref{eq:electroneutrality}) then
takes two equivalent expressions
\begin{equation} \label{eq:rhoelectroneutrality}
q \int_0^{\infty} dz \rho(z) = \sigma , \qquad
\int_0^{\infty} d\widetilde{z} \widetilde{\rho}(\widetilde{z}) = 1 .
\end{equation}
The contact-value theorem for planar wall surfaces \cite{Henderson78} relates
the total contact density of particles to the surface charge density
on the wall and the bulk pressure of the fluid $P$.
For 3D systems of identical particles, it reads
\begin{equation} \label{eq:contacttheorem1}
\beta P = \rho(0) - 2\pi \ell_{\rm B}\sigma^2 .
\end{equation}
Since in the present case of a single isolated double layer, the
pressure vanishes, 
\begin{equation} \label{eq:rhocontacttheorem}
\rho(0) = 2\pi \ell_{\rm B}\sigma^2 , \qquad \widetilde{\rho}(0) = 1 ,
\end{equation}
that can be viewed as a constraint that any reasonable theory
should fulfill.

\subsection{The Virial Strong Coupling approach}

\begin{figure}[htb]
\begin{center}
\includegraphics[width=0.45\textwidth,clip]{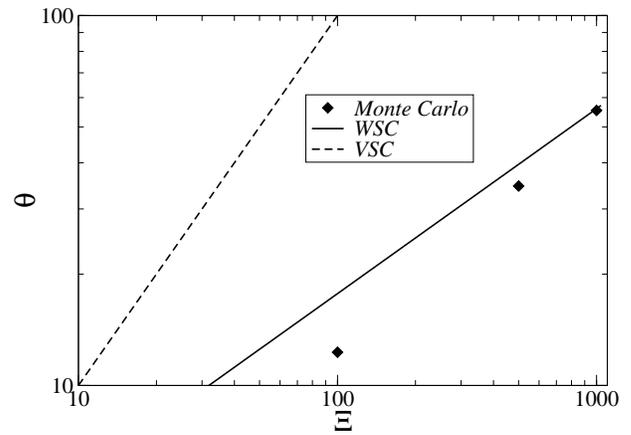}
\caption{The fitting parameter $\theta$, defined by 
Eq. (\ref{eq:theta1}), vs. the coupling constant $\Xi$
for one-plate geometry. 
The MC values reported in Ref. \cite{Moreira00} are shown with filled diamonds,
the original prediction $\theta=\Xi$ of the VSC theory with the dashed line; 
the solid curve is for our WSC prediction, given by Eq. (\ref{eq:prof_theta}).}
\label{fig:prof_theta}
\end{center}
\end{figure}

With our choice of reduced units, the exact density profile
is a function of two variables only: $\widetilde{\rho}(\widetilde{z},\Xi)$.
It is well behaved when $\Xi \to \infty$, which is nevertheless
a limit where in unscaled variables, all counterions stick 
to the plate, forming the Wigner crystal 
($\rho(z,\Xi) \propto \delta(z)$ for $\Xi\to\infty$).
The purpose of the present discussion is to resolve the structure
of the double-layer at large but finite $\Xi$.
According to the VSC method \cite{Moreira00,Netz01}, 
the density profile of counter-ions can be formally expanded in 
the SC regime as a power series in $1/\Xi$:
\begin{equation} \label{eq:virialSC1}
\widetilde{\rho}(\widetilde{z},\Xi) = \widetilde{\rho}_0(\widetilde{z}) +
\frac{1}{\Xi} \widetilde{\rho}_1(\widetilde{z}) 
+ {\cal O}\left( \frac{1}{\Xi^2} \right) , 
\end{equation}
where
\begin{equation} \label{eq:virialSC2}
\widetilde{\rho}_0(\widetilde{z}) = e^{-\widetilde{z}} , \qquad
\widetilde{\rho}_1(\widetilde{z}) = 
e^{-\widetilde{z}} \left( \frac{\widetilde{z}^2}{2} - \widetilde{z} \right) . 
\end{equation}
The leading term $\widetilde{\rho}_0(\widetilde{z})$, which comes from 
the single-particle picture of counter-ions in the linear surface-charge 
potential, is in agreement with the MC simulations \cite{Moreira00}.
Indeed, for large $\Xi$, the particles' excursion perpendicular to the 
plane, which is always quantified by $\mu$, is much smaller than the
lateral spacing between ions (denoted $a$ below) \cite{Netz01}.
As a consequence, these ions experience
the potential of the bare plate, while the interactions with other
ions become negligible by symmetry. 
On the other hand, the MC simulations indicate that the sub-leading term 
$\widetilde{\rho}_1(\widetilde{z})$ has the expected functional form 
(for sufficiently large coupling $\Xi>10$), but the prefactor 
$1/\Xi$ is incorrect.
On the basis of the prediction (\ref{eq:virialSC1}), the MC data were fitted 
in \cite{Moreira00}
by using the formula
\begin{equation} \label{eq:theta1}
\widetilde{\rho}(\widetilde{z},\Xi) - \widetilde{\rho}_0(\widetilde{z}) =
\frac{1}{\theta} \widetilde{\rho}_1(\widetilde{z}) , 
\end{equation}
where $\widetilde{\rho}(\widetilde{z},\Xi)$ is the density profile obtained from
MC simulations and $\theta$ is treated as a fitting parameter.
According to the VSC result (\ref{eq:virialSC1}), $\theta$
should be given by $\theta=\Xi$ plus next-leading corrections.
As is seen in the log-log plot of Fig. \ref{fig:prof_theta}, 
the numerically obtained values of $\theta$ 
are much smaller than $\Xi$, and the difference between $\theta$ and $\Xi$ 
even grows with increasing the coupling constant.

\subsection{The Wigner Strong Coupling expansion}

\begin{figure}[htb]
\begin{center}
\includegraphics[width=0.25\textwidth,clip]{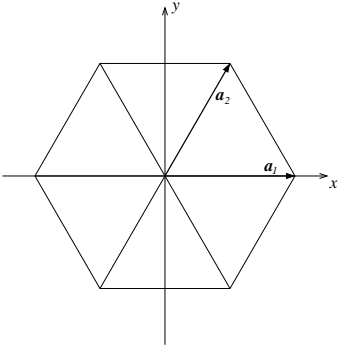}
\caption{Hexagonal structure of the 2D Wigner crystal: 
$\bm{a}_1$ and $\bm{a}_2$ are the primitive translation vectors.}
\label{fig:hexagonal} 
\end{center}
\end{figure}

Our approach is based on the fact that in the asymptotic ground-state limit
$\Xi\to\infty$, all counter-ions collapse on the charged surface $z=0$,
forming a 2D Wigner crystal \cite{Shklovskii,Levin02}.
It is well known \cite{Bonsall77} that the lowest ground-state energy for 
the 2D Wigner crystal is provided by the hexagonal (equilateral triangular)
lattice.
Each point of this lattice has 6 nearest neighbors forming a hexagon, 
see Fig. \ref{fig:hexagonal}.
The 2D lattice points are indexed by 
$\{ j=(j_1,j_2) \}$, where $j_1$ and $j_2$ are any two 
integers (positive, negative or zero):
\begin{equation}
{\bf R}_j = (R_j^x,R_j^y) = j_1 \bm{a}_1 + j_2 \bm{a}_2 ,
\end{equation} 
where
\begin{equation}
\bm{a}_1 = a (1,0) , \qquad 
\bm{a}_2 = a \left( \frac{1}{2},\frac{\sqrt{3}}{2} \right)
\end{equation}
are the primitive translation vectors of the Bravais lattice and $a$ is
the lattice spacing. 
Since at each vertex, there is just one particle, we can identify
$j$ with particle labels, $j=1,\ldots,N$ ($N\to\infty$).
There are two triangles per vertex, so the condition of
global electroneutrality (\ref{eq:electroneutrality}) requires that
\begin{equation} \label{eq:defa}
\frac{q}{\sigma} = \frac{\sqrt{3}}{2} a^2.
\end{equation} 
Note that in the large-$\Xi$ limit, the lateral distance between
the nearest-neighbor counter-ions in the Wigner crystal $a$ is much
larger than the characteristic length $\mu$ in the perpendicular
$z$-direction, $a/\mu\propto \sqrt{\Xi}\gg 1$.
As invoked above, this very feature explains why a single
particle picture provides the leading order term in a SC expansion,
so that the two different approaches discussed here (VSC and WSC)
coincide to leading order. The same remark holds for the two plates
problem that will be addressed in section \ref{sec:twoplates}.
It should be emphasized though that this coincidence of leading
orders is specific to the planar geometry.
The $z$-coordinate of each particle in the ground state is zero, $Z_j=0$.

We denote the ground-state energy of the counter-ions on the Wigner
lattice plus the homogeneous surface-charge density $\sigma e$ by $E_0$.
For $\Xi$ large but not infinite, the fluctuations of ions around their 
lattice positions, in all three spatial directions, begin to play a role.
Let us first shift one of the particles, say $j=1$, from its Wigner lattice
position $({\bf R}_1,Z_1=0)$ by a small vector $\delta{\bf r} = (x,y,z)$
($\vert \delta{\bf r}\vert\ll a$) and look for the corresponding change
in the total energy $\delta E = E - E_0\ge 0$.
The first contribution to $\delta E$ comes from the interaction of 
the shifted counter-ion with the potential induced by the homogeneous
surface charge density:
\begin{equation} \label{eq:firstcontribution}
\delta E^{(1)}(z) = \frac{2\pi q e^2 \sigma}{\varepsilon} z . 
\end{equation}
The second contribution to $\delta E$ comes from the interaction of
the shifted particle $1$ with all other particles $j\ne 1$ on the
2D hexagonal lattice:
\begin{eqnarray}
\delta E^{(2)}(x,y,z) = \phantom{aaaaaaaaaaaaaaaaaaaaaaaaaaaaaaaa} \nonumber \\
\frac{(qe)^2}{\varepsilon} \sum_{j\ne 1} \left[ 
\frac{1}{\sqrt{(R_{1j}^x+x)^2+(R_{1j}^y+y)^2+z^2}} - \frac{1}{R_{1j}} \right] ,
\nonumber \\
\end{eqnarray}
where ${\bf R}_{1j} = (R_{1j}^x,R_{1j}^y) = {\bf R}_1 - {\bf R}_j$ and 
$R_{1j} = \vert {\bf R}_{1j} \vert$. 
Rescaling the lattice positions by $a$ and taking into account the 
inequalities $x/a,y/a,z/a\ll 1$, this expression can be expanded as 
an infinite series in powers of $x/a$, $y/a$ and $z/a$ by using the formula
\begin{equation} \label{eq:texpansion}
\frac{1}{\sqrt{1+t}} = 1 - \frac{1}{2} t + \frac{3}{8} t^2 
- \frac{5}{16} t^3 + \cdots , \quad t\ll 1 . 
\end{equation}
Up to harmonic terms, the expansion reads
\begin{equation} \label{eq:harmonic}
\delta E^{(2)}(x,y,z) = \frac{(qe)^2}{2 \varepsilon a^3} C_3 
\left[ \frac{1}{2}(x^2+y^2) - z^2 \right] .
\end{equation}
Here, $C_3$ is the special $s=3$ case of dimensionless hexagonal 
lattice sums
\begin{equation}
C_s = \sum_{j\ne 1} \frac{1}{(R_{1j}/a)^s} ,
\end{equation}
which can be expressed from the general theory \cite{Zucker} as
\begin{eqnarray}
C_3 & = & \sum_{j,k=-\infty\atop (j,k)\ne (0,0)}^{\infty} 
\frac{1}{(j^2 + j k + k^2)^{3/2}} \nonumber \\
& = & \frac{2}{\sqrt{3}} \zeta\left(\frac{3}{2}\right) \left[
\zeta\left( \frac{3}{2},\frac{1}{3}\right) -
\zeta\left( \frac{3}{2},\frac{2}{3}\right) \right] \label{eq.C3}
\end{eqnarray}
with $\zeta(z,q) = \sum_{n=0}^{\infty} 1/(q+n)^z$ the generalized
Riemann zeta function and $\zeta(z) \equiv \zeta(z,1)$
(this function should not be confused with the parameter
$\zeta$, appearing without arguments below after Eq. (\ref{eq:defzeta}), 
that will measure the asymmetry between two charged plates). Explicitly,
$C_3=11.034\ldots$.
The absence of the linear $x$, $y$ terms and of the mixed $xy$ term
in (\ref{eq:harmonic}) is caused by the fact that every lattice point 
is at a center of inversion.
The invariance of the hexagonal lattice with respect to the rotation 
around any point by the angle $\pi/3$ implies the lattice sum equalities
\begin{eqnarray}
\sum_{j\ne 1} f(R_{1j}) \left( R_{1j}^x \right)^2 & = &
\sum_{j\ne 1} f(R_{1j}) \left( R_{1j}^y \right)^2 \nonumber \\
& = & \frac{1}{2} \sum_{j\ne 1} f(R_{1j}) R_{1j}^2 ,
\end{eqnarray}
which were also used in the derivation of (\ref{eq:harmonic}).
Note that the $x^2$ and $y^2$ harmonic terms in Eq. (\ref{eq:harmonic}) 
have positive signs which is consistent with the stability of the Wigner
crystal in the $(x,y)$ plane.
On the other hand, the minus sign of the $z^2$ term does not represent
any stability problem due to the presence of the positive linear contribution
in (\ref{eq:firstcontribution}), which is dominant for small $z$-distances.
The total energy change is given by 
$\delta E(x,y,z) = \delta E^{(1)}(z) + \delta E^{(2)}(x,y,z)$.
Finally, let us write down the $z$-dependent part of the dimensionless energy
shift
$-\beta \delta E$, with $z$ expressed in units of $\mu$:
\begin{equation}
-\beta \delta E(0,0,\mu\widetilde{z}) \sim -\widetilde{z} 
+ \frac{\alpha^3}{2} \frac{C_3}{\sqrt{\Xi}} \widetilde{z}^2 , \quad
\alpha = \frac{3^{1/4}}{2\sqrt{\pi}} .
\end{equation}
We see that in the limit $\Xi\to\infty$, as advocated above, 
the two-body interaction term of 
the shifted ion with all other ions on the Wigner crystal is of order
$1/\sqrt{\Xi}$ and therefore negligible in comparison with the one-body 
potential term $-\widetilde{z}$ due to the surface charge density.
This leading single-particle picture is common to both VSC and WSC approaches.
As concerns the two-body interaction terms $\widetilde{z}^p$ of higher 
orders ($p=3,4,\ldots$), their coefficients are proportional to
$q^2 \ell_{\rm B} \mu^p/a^{p+1} \propto 1/\Xi^{(p-1)/2}$.
The present scheme thus represents a systematic basis for an expansion
in powers of $1/\sqrt{\Xi}$.   

The generalization of the above formalism to independent shifts of all 
particles from their lattice positions is straightforward.
Let us shift every particle $j=1,2,\ldots,N$ from its lattice position
$({\bf R}_j,Z_j=0)$ by a small vector $\delta {\bf r}_j=(x_j,y_j,z_j)$
($\vert \delta {\bf r}_j \vert\ll a$) and study the corresponding energy
change $\delta E$.
As before, the first (one-body) contribution to $\delta E$ is given by 
\begin{equation} 
-\beta \delta E^{(1)}(\{\mu\widetilde{z}_j\}) = - \sum_{j=1}^N \widetilde{z}_j . 
\end{equation}
The second (two-body) contribution to $\delta E$ is expressible as
\begin{eqnarray}
\delta E^{(2)}(\{x_j\},\{y_j\},\{z_j\}) = \phantom{aaaaaaaaaaaaaaaaaa} 
\nonumber \\
\frac{(qe)^2}{2 \varepsilon} \sum_{j,k=1\atop (j\ne k)}^N \frac{1}{R_{jk}}
\left[ \frac{1}{\sqrt{1+\mu_{jk}+\nu_{jk}}} - 1 \right] ,
\end{eqnarray} 
where the dimensionless $\mu_{jk}$ and $\nu_{jk}$ involve the particle 
coordinates along and perpendicular to the Wigner crystal, respectively:
\begin{eqnarray}
\mu_{jk} & = & 2(x_j-x_k) \frac{R_{jk}^x}{R_{jk}^2} + 
2(y_j-y_k) \frac{R_{jk}^y}{R_{jk}^2} \nonumber \\
& & + \frac{1}{R_{jk}^2} \left[ (x_j-x_k)^2 + (y_j-y_k)^2 \right] , \\
\nu_{jk} & = & \frac{1}{R_{jk}^2} (z_j-z_k)^2 .
\end{eqnarray}
Performing the expansion of type (\ref{eq:texpansion}) in small $\mu_{jk}$ 
and $\nu_{jk}$, we end up with
\begin{equation}
-\beta \delta E^{(2)}(\{x_j\},\{y_j\},\{z_j\}) = S_z + S_W + S_{z,W} ,
\end{equation}
where
\begin{equation} \label{Sz}
S_z = \frac{q^2 \ell_{\rm B}}{2} \sum_{j,k=1\atop (j\ne k)}^N \frac{1}{R_{jk}}
\left( \frac{1}{2} \nu_{jk} - \frac{3}{8} \nu_{jk}^2 + \cdots \right)
\end{equation}
contains particle shifts exclusively in the $z$ direction,
\begin{equation} \label{Sw}
S_W = \frac{q^2 \ell_{\rm B}}{2} \sum_{j,k=1\atop (j\ne k)}^N \frac{1}{R_{jk}}
\left( \frac{1}{2} \mu_{jk} - \frac{3}{8} \mu_{jk}^2 + \cdots \right)
\end{equation}
contains particle shifts exclusively in the $(x,y)$ Wigner plane and
\begin{eqnarray} \label{Szw}
S_{z,W} & = & \frac{q^2 \ell_{\rm B}}{2} \sum_{j,k=1\atop (j\ne k)}^N \frac{1}{R_{jk}}
\left[ - \frac{3}{4} \mu_{jk} \nu_{jk} \right. \nonumber \\ & & \left.
+ \frac{15}{16} 
\left( \mu_{jk}^2 \nu_{jk} + \mu_{jk} \nu_{jk}^2 \right) + \cdots \right]
\end{eqnarray}
mixes particle shifts along the $z$ direction with those along 
the $(x,y)$ plane.

We are interested in the particle density profile defined by
$\rho({\bf r}) = \langle \sum_{j=1}^N \delta({\bf r}-{\bf r}_j) \rangle$,
where $\langle \cdots \rangle$ means thermal equilibrium average over
the Boltzmann weight $\exp(-\beta\delta E)$ with
\begin{eqnarray} \label{dE}
-\beta \delta E & = & -\beta \delta E^{(1)} -\beta \delta E^{(2)} \nonumber \\
& = & -\sum_{j=1}^N \widetilde{z}_j + S_z + S_W +S_{z,W} .
\end{eqnarray}
The ground-state energy $E_0$ is a quantity which is independent of the particle 
coordinate shifts and as such disappears for the statistical averages.
The system is translationally invariant in the $(x,y)$ plane, so that
the particle density is only $z$-dependent, $\rho({\bf r})=\rho(z)$.
We shall consider separately in (\ref{dE}) the terms containing exclusively
particle shifts in $z$ direction, transversal to the wall, and those
which involve longitudinal particle shifts along the Wigner $(x,y)$ plane.   

\subsection{Contribution of transversal particle shifts}
Let us forget for a while the terms $S_W$ and $S_{z,W}$ in (\ref{dE}) 
and consider only the particle $z$-shifts in the ``most relevant'' $S_z$,
\begin{equation} \label{eq:deltae}
-\beta\delta E = - \sum_{j=1}^N \widetilde{z}_j + S_z . 
\end{equation}
Expressing $z$ in units of $\mu$, $S_z$ in Eq. (\ref{Sz}) can be written as 
an infinite series in powers of $1/\sqrt{\Xi}$, the first terms of which read
\begin{eqnarray} \label{eq:Sz}
S_z & = & \frac{\alpha^3}{4\sqrt{\Xi}} \sum_{j,k=1\atop (j\ne k)}^N 
\frac{1}{(R_{jk}/a)^3} (\widetilde{z}_j-\widetilde{z}_k)^2 \nonumber \\
& & - \frac{3 \alpha^5}{16 \,\Xi^{3/2}} \sum_{j,k=1\atop (j\ne k)}^N 
\frac{1}{(R_{jk}/a)^5} (\widetilde{z}_j-\widetilde{z}_k)^4 + \cdots .
\end{eqnarray}  
In the limit $\Xi\to\infty$, 
$S_z$ is a perturbation with respect to the one-body part in (\ref{eq:deltae}).

To obtain the particle density, we add to the one-body potential $\widetilde{z}$ 
an auxiliary (generating or source) potential $\beta u({\bf r})$,
which will be set to $0$ at the end of calculations.
The partition function of our $N$-particle system
\begin{equation} \label{eq:partition}
Z_N[w] = \frac{1}{N!} \int_{\Lambda} \prod_{i=1}^N \left[ d{\bf r}_i 
w({\bf r}_i) e^{-\widetilde{z}_i} \right] \exp(S_z) 
\end{equation}
thereby becomes a functional of the generating Boltzmann weight
$w({\bf r})=\exp[-\beta u({\bf r})]$.
The particle density at point ${\bf r}$ is obtained as the
functional derivative
\begin{equation} \label{eq:density}
\rho({\bf r}) = \frac{\delta}{\delta w({\bf r})} \ln Z_N[w] 
\Big\vert_{w({\bf r})=1} ,
\end{equation}
which is of course a function of $\Xi$, in addition to ${\bf r}$.
To treat $S_z$ as the perturbation, we define the $S_z=0$ counterpart
of the partition function (\ref{eq:partition})
\begin{eqnarray}
Z_N^{(0)}[w] & = & \frac{1}{N!} \int_{\Lambda} \prod_{i=1}^N \left[ d{\bf r}_i 
w({\bf r}_i) e^{-\widetilde{z}_i} \right] \nonumber \\ & = &  \frac{1}{N!} 
\left[ \int_{\Lambda} d{\bf r} w({\bf r}) e^{-\widetilde{z}} \right]^N , 
\end{eqnarray} 
which corresponds to non-interacting particles in an external potential. 
It is clear that
\begin{equation}
\ln \left( \frac{Z_N[w]}{Z_N^{(0)}[w]} \right) = 
\ln \langle \exp(S_z) \rangle_0 ,
\end{equation}
where $\langle \cdots \rangle_0$ denotes the averaging over the system 
of non-interacting particles defined by $Z_N^{(0)}$.
We are left with the cumulant expansion of $\ln\langle\exp(S_z)\rangle_0$:
\begin{eqnarray}
\ln \langle \exp(S_z) \rangle_0 & = & \sum_{n=1}^{\infty} \frac{1}{n!}
\langle S_z^n \rangle_0^{(c)} \nonumber \\
& = & \langle S_z \rangle_0 + \frac{1}{2} \left(
\langle S_z^2 \rangle_0 - \langle S_z \rangle_0^2 \right) + \cdots . 
\end{eqnarray} 
An important property of the cumulant expansion is that if
$\langle S_z \rangle_0$ is an extensive (proportional to $N$) quantity, 
the higher-order terms will also be.
In other words, the contributions of $N^2$, $N^3$, etc. orders
will cancel with each other.
We conclude that
\begin{equation} \label{eq:lnZ}
\ln Z_N[w] = \ln Z_N^{(0)}[w] + \langle S_z\rangle_0 + \frac{1}{2} 
\left( \langle S_z^2 \rangle_0 - \langle S_z \rangle_0^2 \right) + \cdots . 
\end{equation}
The particle density results from the substitution of this expansion 
into (\ref{eq:density}), and the subsequent application of the functional 
derivative with respect to $w({\bf r})$, taken at $w({\bf r})=1$.

The leading SC behavior of the particle density stems from $\ln Z_N^{(0)}[w]$.
Since
\begin{eqnarray}
\frac{\delta}{\delta w({\bf r})} \ln Z_N^{(0)}[w] \Big\vert_{w({\bf r})=1}
& = & \frac{N e^{-\widetilde{z}}}{\int_{\Lambda} d{\bf r} e^{-\widetilde{z}}}
= \frac{N}{\vert\Sigma\vert\mu} e^{-\widetilde{z}} \nonumber \\ 
& = & (2\pi\ell_{\rm B}\sigma^2) e^{-\widetilde{z}} 
\end{eqnarray}
we have $\widetilde{\rho}_0(\widetilde{z}) \sim e^{-\widetilde{z}}$,
which coincides with the leading VSC term presented in (\ref{eq:virialSC2}).

\begin{figure}[htb]
\begin{center}
\includegraphics[width=0.45\textwidth,clip]{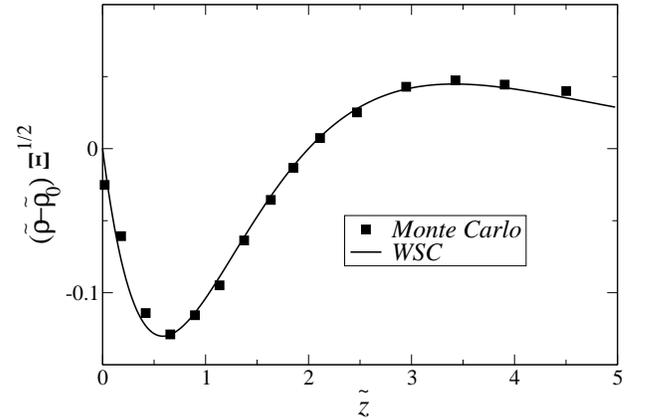}
\caption{Single charged wall: Comparison between the
rescaled analytical first correction to the strong coupling profile from 
Eq. (\ref{eq:leadingpluscorrection}) (solid curve) and the MC results 
of Ref. \cite{Moreira00} (filled squares). Here, $\Xi=10^3$ 
and $\widetilde\rho_0(z)$ denotes
the leading order term $\exp(-\widetilde z)$, that is subtracted from the 
numerical data to probe the correction.} 
\label{fig:prof} 
\end{center}
\end{figure}

The first correction to the density profile stems from 
$\langle S_z \rangle_0$, namely from the first term in 
the series representation of $S_z$ (\ref{eq:Sz}): 
\begin{equation} \label{eq:firstcorrection}
\langle S_z \rangle_0 \sim \frac{\alpha^3}{4\sqrt{\Xi}}
\sum_{j,k=1\atop (j\ne k)}^N \frac{1}{(R_{jk}/a)^3} \langle \left( 
\widetilde{z}_j^2+\widetilde{z}_k^2-2\widetilde{z}_j\widetilde{z}_k \right) \rangle_0 .
\end{equation}
A useful property of the averaging $\langle \cdots \rangle_0$ is
its independence on the particle (lattice site) index, e.g. for
$p=1,2,\ldots$ we have
\begin{eqnarray}
\langle \widetilde{z}_j^p \rangle_0 & = & 
\frac{\int_{\Lambda} \prod_{i=1}^N \left[ d{\bf r}_i w({\bf r}_i) e^{-\widetilde{z}_i} 
\right] \widetilde{z}_j^p}{\int_{\Lambda} \prod_{i=1}^N \left[ d{\bf r}_i w({\bf r}_i) 
e^{-\widetilde{z}_i} \right]} \nonumber \\ & = &
\frac{\int_{\Lambda}d{\bf r} w({\bf r}) e^{-\widetilde{z}} \widetilde{z}^p}{
\int_{\Lambda}d{\bf r} w({\bf r}) e^{-\widetilde{z}}} \equiv [\widetilde{z}^p]_0 .
\end{eqnarray}
Simultaneously, due to the absence of interactions in 
$\langle \cdots \rangle_0$, correlation functions of particles
decouple themselves, e.g. 
$\langle \widetilde{z}_j \widetilde{z}_k \rangle_0 = [\widetilde{z}]_0^2$ for $j\ne k$.
Thus, the relation (\ref{eq:firstcorrection}) becomes
\begin{equation}
\langle S_z \rangle_0 \sim \frac{\alpha^3}{2\sqrt{\Xi}} N C_3
\left( [\widetilde{z}^2]_0 - [\widetilde{z}]_0^2 \right) .
\end{equation}
It is easy to show that
\begin{equation}
\frac{\delta}{\delta w({\bf r})} [\widetilde{z}^p]_0 \Big\vert_{w({\bf r})=1}
= \frac{1}{\vert\Sigma\vert\mu} e^{-\widetilde{z}} \left( \widetilde{z}^p - p! \right) ,
\end{equation}
where we used the equality $[\widetilde{z}^p]_0\vert_{w({\bf r})=1} = p!$.
The formula for the density profile, in the leading order plus the first
correction, then reads
\begin{equation} \label{eq:leadingpluscorrection}
\widetilde{\rho}(\widetilde{z},\Xi) = e^{-\widetilde{z}} + \frac{3^{3/4}}{8\pi^{3/2}}
\frac{C_3}{\sqrt{\Xi}} e^{-\widetilde{z}} 
\left( \frac{\widetilde{z}^2}{2} - \widetilde{z}  \right)
+ {\cal O}\left( \frac{1}{\Xi} \right) .
\end{equation}
Note that the electroneutrality (\ref{eq:rhoelectroneutrality}) and
the contact theorem (\ref{eq:rhocontacttheorem}) are satisfied by 
this density profile.
In Fig. \ref{fig:prof}, we compare the appropriately rescaled first correction
to the leading SC profile obtained in (\ref{eq:leadingpluscorrection})
(solid curve) with MC data \cite{Moreira00} at $\Xi=10^3$ (filled squares). 
The agreement is excellent. 
On the other hand, the VSC prediction is off by a factor $1000^{1/2}$.

Comparing our WSC result (\ref{eq:leadingpluscorrection}) with the
VSC Eqs. (\ref{eq:virialSC1}) and (\ref{eq:virialSC2})
we see that the first corrections have the same functional dependence in 
$\widetilde{z}$, but different prefactors.
In terms of the fitting parameter $\theta$ introduced in (\ref{eq:theta1}),
the VSC estimate $\theta=\Xi$ is compared with the present value
\begin{equation} \label{eq:prof_theta}
\theta = \frac{8\pi^{3/2}}{3^{3/4}} \frac{1}{C_3} \sqrt{\Xi}
= 1.771\ldots \sqrt{\Xi} .
\end{equation}
As is seen from Fig. \ref{fig:prof_theta}, this formula (solid curve) 
is in full agreement with the data of MC simulations (filled diamonds).

In the series representation of $S_z$ (\ref{eq:Sz}), the first term
is of order $\Xi^{-1/2}$ and the second one is of order $\Xi^{-3/2}$.
In view of (\ref{eq:lnZ}), the second correction to the density profile 
stems from $(\langle S_z^2 \rangle_0-\langle S_z \rangle_0^2)/2$ with
$S_z$ represented by its first term, and not from $\langle S_z \rangle_0$ 
with $S_z$ represented by its second term.
Let us analyze in detail the average
\begin{eqnarray}
\langle S_z^2 \rangle_0 & \sim & \left( \frac{\alpha^3}{4\sqrt{\Xi}} \right)^2
\sum_{(j\ne k)} \frac{1}{(R_{jk}/a)^3} \sum_{(m\ne n)} \frac{1}{(R_{mn}/a)^3}
\nonumber \\ & & \times
\frac{\int_{\Lambda} \prod_{i=1}^N \left[ d{\bf r}_i w({\bf r}_i) e^{-\widetilde{z}_i} 
\right] (\widetilde{z}_j-\widetilde{z}_k)^2 (\widetilde{z}_m-\widetilde{z}_n)^2}{
\int_{\Lambda} \prod_{i=1}^N \left[ d{\bf r}_i w({\bf r}_i) e^{-\widetilde{z}_i} 
\right]} . \nonumber \\ & &
\end{eqnarray}
For a fixed pair of site indices $(j\ne k)$, there exist seven topologically 
different possibilities for the pair $(m\ne n)$:
$$
\left.
\begin{array}{cc}
m=j, & n=k; \cr
n=j, & m=k;
\end{array}
\right\} \qquad \mbox{factor $2$}
$$ 
$$
\left.
\begin{array}{cc}
m=j, & n\ne j,k; \cr
n=j, & m\ne j,k; \cr
m=k, & n\ne j,k; \cr
n=k, & m\ne j,k; \cr
\end{array}
\right\} \qquad \mbox{factor $4$}
$$ 
$$
m\ne j,k, \quad n\ne j,k,m. \} \qquad \mbox{factor $1$} 
$$
Here, respecting the properties of the averaging $\langle\cdots\rangle_0$, 
those possibilities which lead to the same result are grouped together. 
After simple algebra, we find that
\begin{eqnarray}
\langle S_z^2 \rangle_0 & \sim & 
\frac{\alpha^6}{4 \Xi} \Big\{ N C_3^2 \left( [\widetilde{z}^4]_0 
- 4 [\widetilde{z}^3]_0 [\widetilde{z}]_0 + 3 [\widetilde{z}^2]_0^2 \right) 
\nonumber \\ & & + [(NC_3)^2-4NC_3^2+2NC_6]
\left( [\widetilde{z}^2]_0 - [\widetilde{z}]_0^2 \right)^2 \Big\}. 
\nonumber \\ & &
\end{eqnarray}
The ``undesirable'' disconnected term of order $N^2$ is cancelled by 
the subtraction of $\langle S_z \rangle_0^2$.
After performing the functional derivatives with respect to 
$w({\bf r})$, taken at $w({\bf r})=1$, we end up with the next
correction to the profile (\ref{eq:leadingpluscorrection}) of the form
\begin{equation}
\frac{3^{3/2}}{64 \pi^3} \frac{1}{\Xi} e^{-\widetilde{z}} \left[
C_3^2 \left( \frac{\widetilde{z}^4}{8} - \frac{\widetilde{z}^3}{2} +
\frac{\widetilde{z}^2}{2} - \widetilde{z} \right) +
C_6 \left( \frac{\widetilde{z}^2}{2} - \widetilde{z} \right) \right] .
\end{equation}
Note that this correction does not break the electroneutrality
condition (\ref{eq:rhoelectroneutrality}) nor the contact theorem
(\ref{eq:rhocontacttheorem}).

\subsection{Contribution of longitudinal and mixed particle shifts}
Now we consider in (\ref{dE}) also the term $S_W$ with purely longitudinal 
particle shifts in the Wigner plane and the term $S_{z,W}$ with 
mixed transversal and longitudinal shifts.
Denoting particle shifts in the infinite Wigner plane as 
${\bf u}_j=(x_j,y_j)$, these terms possess the important 
translational symmetry:
\begin{eqnarray} 
S_W(\{ {\bf u}_j \}) & = & S_W(\{ {\bf u}_j+{\bf u} \}) , \nonumber \\
S_{z,W}(\{ {\bf u}_j,z_j \}) & = & S_{z,W}(\{ {\bf u}_j+{\bf u},z_j \}) ,
\label{eq:transl}
\end{eqnarray}
where ${\bf u}$ is any 2D vector.
We first investigate the scaling properties of $S_W$ and $S_{z,W}$.

Let us expand $S_W$ up to quadratic $x,y$-deviations:
\begin{eqnarray} \label{Sw2}
S_W & = & \frac{q^2\ell_{\rm B}}{4a} \sum_{j,k=1\atop (j\ne k)}^N
\frac{(R_{jk}^y/a)^2-2(R_{jk}^x/a)^2}{(R_{jk}/a)^5} 
\left( \frac{x_j-x_k}{a} \right)^2 \nonumber \\
& & + \frac{q^2\ell_{\rm B}}{4a} \sum_{j,k=1\atop (j\ne k)}^N
\frac{(R_{jk}^x/a)^2-2(R_{jk}^y/a)^2}{(R_{jk}/a)^5} 
\left( \frac{y_j-y_k}{a} \right)^2 \nonumber \\
& & - \frac{3q^2\ell_{\rm B}}{2a} \sum_{j,k=1\atop (j\ne k)}^N
\frac{(R_{jk}^x R_{jk}^y)/a^2}{(R_{jk}/a)^5} \frac{(x_j-x_k)(y_j-y_k)}{a^2}  
\nonumber \\ & & + \cdots .
\end{eqnarray}
Terms linear in $(x_j-x_k)/a$ and $(y_j-y_k)/a$ vanish because every
point of the hexagonal Wigner crystal is a center of inversion. 
We saw that in the $z$ direction the relevant length scale is determined
by the Gouy-Chapman length $\mu$: Rescaling the $z$ coordinates by $\mu$, 
the (leading) linear potential term $\widetilde{z}$ is independent of 
the coupling constant $\Xi$ while the next terms are proportional to
inverse powers of $\sqrt{\Xi}$ and therefore vanish in the SC limit.
The natural length scale in the Wigner $(x,y)$ plane is the lattice spacing 
$a$, but this is not the relevant scale in statistical averages.
The relevant length $\lambda$ is determined by the requirement that 
the rescaling of coordinates $x_j=\lambda X_j$ and $y_j=\lambda Y_j$ 
in (\ref{Sw2}) leads to a dimensionless and $\Xi$-independent (leading) 
quadratic term.
Since $q^2\ell_{\rm B}/a \propto \sqrt{\Xi}$, we have
\begin{equation}
\frac{\lambda}{a} \propto \frac{1}{\Xi^{1/4}} , \qquad
\frac{\lambda}{\mu} \propto \Xi^{1/4} 
\end{equation} 
(the numerical prefactors are unimportant),
i.e. the relevant scale is ``in between'' $\mu$ and $a$.
The higher-order terms in $S_W$, which contain the deviations $(x_j-x_k)$ 
and $(y_j-y_k)$ in powers $p=3,4,\ldots$, scale like $1/\Xi^{(p-2)/4}$
and therefore vanish in the limit $\Xi\to\infty$.

Let us now consider the leading expansion terms of the mixed 
quantity $S_{z,W}$:
\begin{eqnarray}
S_{z,W} & = & - \frac{3 q^2\ell_{\rm B}}{4a} \sum_{j,k=1\atop (j\ne k)}^N
\frac{[(z_j-z_k)/a]^2}{(R_{jk}/a)^5} \nonumber \\ & & \times
\left[ \frac{R_{jk}^x}{a} \left( \frac{x_j-x_k}{a} \right) 
+ \frac{R_{jk}^y}{a} \left( \frac{y_j-y_k}{a} \right) \right] \nonumber \\
& & + \frac{3 q^2\ell_{\rm B}}{8a} \sum_{j,k=1\atop (j\ne k)}^N
\frac{[(z_j-z_k)/a]^2}{(R_{jk}/a)^7} \nonumber \\ & & \times
\Bigg\{ \left[ 4 \left( \frac{R_{jk}^x}{a} \right)^2 - 
\left( \frac{R_{jk}^y}{a} \right)^2  \right] \left( \frac{x_j-x_k}{a} \right)^2 
\nonumber \\ & & + \left[ 4 \left( \frac{R_{jk}^y}{a} \right)^2 - 
\left( \frac{R_{jk}^x}{a} \right)^2  \right] \left( \frac{y_j-y_k}{a} \right)^2 
\nonumber \\ & & + 10 \frac{R_{jk}^x}{a} \frac{R_{jk}^y}{a} 
\left( \frac{x_j-x_k}{a} \right) \left( \frac{y_j-y_k}{a} \right) 
\Bigg\} \nonumber \\ & & + \cdots .
\end{eqnarray}
Rescaling the particle coordinates as follows $z_j=\mu \widetilde{z}_j$,
$x_j=\lambda X_j$, $y_j=\lambda Y_j$, the first term is of order
$1/\Xi^{3/4}$ and the second one is of order $1/\Xi$.

To obtain the density profile, one proceeds in analogy with the previous case
of transversal vibrations.
We introduce the partition function of our $N$-particle system
\begin{equation} \label{eq:partition2}
Z_N[w] = \frac{1}{N!} \int_{\Lambda} \prod_{j=1}^N \left[ d{\bf r}_i 
w({\bf r}_i) e^{-\widetilde{z}_i} \right] e^{S_W} e^{S_z+S_{z,W}} 
\end{equation}
with the generating Boltzmann weight $w({\bf r})$.
We take as the unperturbed system the one with one-body potentials
$-\widetilde{z}_i$ in $z$ direction and $S_W$ in $(x,y)$ plane, and
treat $S_z+S_{z,W}$ as the perturbation. 
Using the cumulant method, we obtain
\begin{equation}
\ln Z_N[w] = \ln Z_N^{(0)}[w] + \langle S_z\rangle_0 +
\langle S_{z,W}\rangle_0 + \cdots ,
\end{equation}
where $\langle \cdots \rangle_0$ denotes the averaging over the
unperturbed system with the partition function
\begin{equation}
Z_N^{(0)}[w] = \frac{1}{N!} \int_{\Lambda} \prod_{i=1}^N \left[ d{\bf r}_i 
w({\bf r}_i) e^{-\widetilde{z}_i} \right] \exp(S_W) . 
\end{equation} 
The particle density is given by Eq. (\ref{eq:density}).

The additional appearance of $\exp(S_W)$ in the averaging over the
unperturbed system is a complication which can be sometimes removed trivially 
by using the translational invariance of $S_W$ (\ref{eq:transl}).
We shall document this fact on the leading SC behavior of the particle
density at point ${\bf r}=({\bf u},z)$ which stems from $\ln Z_N^{(0)}[w]$:
\begin{eqnarray}
\frac{\delta}{\delta w({\bf r})} \ln Z_N^{(0)}[w] \Big\vert_{w({\bf r})=1}
\phantom{aaaaaaaaaaaaaaaaaa} \nonumber \\ =  \frac{N e^{-\widetilde{z}}}{\mu} 
\frac{\int_{\Sigma}\prod_{i=2}^N d^2 u_i e^{S_W({\bf u}_1={\bf u})}}{
\int_{\Sigma}\prod_{i=1}^N d^2 u_i e^{S_W}} . \label{eq:Z0}
\end{eqnarray}
Since the surface of the plate $\Sigma$ is infinite, we shift
in the denominator the integral variables $i\ne 1$ as follows
${\bf u}_i\to {\bf u}_i+{\bf u}_1-{\bf u}$ which transforms
$S_W\to S_W({\bf u}_1={\bf u})$.
Integrating over ${\bf u}_1$, the ratio of integrals in (\ref{eq:Z0}) 
is ${\bf u}$-independent, and reads $1/\vert\Sigma\vert$.
By this simple technique, it can be shown that the contribution to
the density profile coming from the functional derivative of
$\langle S_z\rangle_0$ is not affected by $S_W$, which decouples from
the $z$-variables.
We remember from the previous part about transversal deviations
that $\langle S_z \rangle_0$ is of order $1/\sqrt{\Xi}$.

The description is a bit more complicated in the case of
\begin{equation}
\langle S_{z,W} \rangle_0 = \frac{\int_{\Lambda} \prod_{i=1}^N \left[ d{\bf r}_i 
w({\bf r}_i) e^{-\widetilde{z}_i} \right] \exp(S_W) S_{z,W}}{
\int_{\Lambda} \prod_{i=1}^N \left[ d{\bf r}_i w({\bf r}_i) 
e^{-\widetilde{z}_i} \right] \exp(S_W)} .
\end{equation}
In the corresponding contribution to the density profile, obtained 
as the functional derivative with respect to $w({\bf r})$ at $w({\bf r})=1$, 
the $z$ and $(x,y)$ subspaces decouple from one another.
The $z$ variables are considered in the rescaled form $\widetilde{z}=z/\mu$.
To perform the integration over the Wigner plane, we rescale
the $(x,y)$ variables to the ones $\lambda(X,Y)$; this ensures that
the quadratic part of $S_W$ is $\Xi$-independent and all higher-order
terms $p=3,4,\ldots$, proportional to $1/\Xi^{(p-2)/4}$, vanish in the
SC limit $\Xi\to\infty$.
Thus the leading dependence on $\Xi$ is given by the scaling factor
of $S_{z,W}$ under the coordinate transformations $z=\mu\widetilde{z}$
and $(x,y)=\lambda(X,Y)$, which was found to be of order $1/\Xi^{3/4}$.
This contribution does not alter the first correction $\propto 1/\sqrt{\Xi}$. 
To calculate explicitly the second correction is a complicated task,
because the quadratic part of $S_W$ in the exponential $\exp(S_W)$
involves all interactions of particles on the Wigner crystal.
The explicit diagonalization of $S_W$ can be done e.g. in the small wave 
vector limit \cite{Bonsall77}.

The fact that the longitudinal vibrations in the plane of the Wigner crystal
have no effect on the leading term and the first correction of the particle
density profile is a general feature of the WSC theory.
In what follows, we shall ignore these degrees of freedom, 
restricting ourselves to the leading term and the first correction,
proportional to $1/\sqrt{\Xi}$.

\section{Parallel plates at small separation}
\label{sec:twoplates}
Next we study the geometry of two parallel plates $\Sigma_1\equiv 1$ and 
$\Sigma_2\equiv 2$ of the same (infinite) surface 
$\vert \Sigma_1\vert = \vert \Sigma_2\vert = \vert\Sigma\vert$, separated by a
distance $d$, see Fig. \ref{fig:geometry}b.
The $z=0$ plate 1 carries the constant surface charge density $\sigma_1 e$, 
while the other plate 2 at $z=d$ is charged by $\sigma_2 e$.
The electric potential between the plates is, up to an irrelevant constant, 
given by
\begin{equation} \label{eq:one-body}
\phi(z) = -\frac{2\pi (\sigma_1-\sigma_2)e}{\varepsilon} z.
\end{equation}
$N$ mobile counter-ions of charge $- q e$ (the valency $q>0$), which are in 
the region between the walls $\Lambda = \left\{ {\bf r},0\le z\le d \right\}$, 
compensate exactly the fixed charge on the plates:
\begin{equation} \label{eq:electro1}
q N = (\sigma_1+\sigma_2) \vert\Sigma\vert .
\end{equation}
Without any loss of generality we can assume $\sigma_1>0$, so that
the asymmetry parameter
\begin{equation}
\zeta = \frac{\sigma_2}{\sigma_1} \ge -1 .
\label{eq:defzeta}
\end{equation}
This parameter should not be confused with the Riemann function
introduced in Eq. (\ref{eq.C3}).
By rescaling appropriately model's parameters, it is sufficient
to consider the interval $-1\le \zeta \le 1$.
The limiting value $\zeta=-1$ corresponds to the trivial case
$\sigma_2=-\sigma_1$ with no counter-ions between the plates.
The symmetric case $\zeta=1$ corresponds to equivalently charged
plates $\sigma_2=\sigma_1$.  Note that in all cases considered,
there is only one type of mobile ion in the interstitial space
$0\leq z \leq d$.

Because of the asymmetry between the surface charges, there exist two
Gouy-Chapman lengths
\begin{equation}
\mu_1 = \frac{1}{2\pi\ell_{\rm B}q \sigma_1} \equiv \mu , \quad
\mu_2 = \frac{1}{2\pi\ell_{\rm B}q \vert\sigma_2\vert} = 
\frac{\mu}{\vert\zeta\vert}.
\end{equation}
Similarly, we can define two different coupling parameters
\begin{equation}
\Xi_1 = \frac{q^2 \ell_{\rm B}}{\mu_1} \equiv \Xi , \quad
\Xi_2 = \frac{q^2 \ell_{\rm B}}{\mu_2} = \vert\zeta\vert \Xi .
\end{equation}
Here, for the ease of comparison, 
we follow the convention of Ref. \cite{Kanduc08}: 
all quantities will be rescaled by their plate 1 counterparts,
i.e. $\widetilde{z}=z/\mu_1$, and
\begin{equation}
\widetilde{\rho}(\widetilde{z}) = \frac{\rho(\mu\widetilde{z})}{2\pi\ell_{\rm B}\sigma_1^2} ,
\quad \widetilde{P} = \frac{\beta P}{2\pi\ell_{\rm B}\sigma_1^2} .
\end{equation}
The reduced density is a function of three arguments: $\widetilde z$,
$\widetilde d$ and $\Xi$ while the reduced pressure depends on two:
$\widetilde d$ and $\Xi$. For notational simplicity, the dependence
on $\widetilde d$ and $\Xi$ will often be implicit in what follows.
Note also that $\widetilde P = \epsilon P / (2 \pi e^2 \sigma_1^2)$, so that 
the rescaling factor required to defined the dimensionless pressure is temperature independent.
This is not the case of the rescaling factor applied to distances, since the Gouy-Chapman
lengths scale as $T$.
The electroneutrality condition (\ref{eq:electro1}) can be written in two
equivalent ways
\begin{equation} \label{eq:electro2}
\int_0^d dz \rho(z) = \frac{\sigma_1+\sigma_2}{q} , \quad
\int_0^{\widetilde{d}} d\widetilde{z} \, \widetilde{\rho}(\widetilde{z}) \, =\,  1+\zeta .
\end{equation}
The contact-value theorem (\ref{eq:contacttheorem1}), considered
at $z=0$ and $z=d$ boundaries, takes two equivalent forms
\begin{equation} \label{eq:contacttheorem2}
\widetilde{P} = \widetilde{\rho}(0) - 1 = \widetilde{\rho}(\widetilde{d}) - \zeta^2 ,
\end{equation}
which provides a strong $d$ and $\Xi$ independent 
constraint for $\widetilde{\rho}(0) -\widetilde{\rho}(d)$.

In the case of oppositely charged surfaces $-1<\zeta\le 0$, the ground state
of the counter-ion system is the same as for the isolated plate 1, i.e.
all $N$ counter-ions collapse on the surface, and create 
the hexagonal Wigner crystal.
For this region of $\zeta$ values, one can easily adapt the WSC technique 
from the one-plate geometry for {\it a priori} 
any distance $d$ between the plates. 

The case of like-charged plates $0<\zeta\le 1$ is more subtle. The ground state
of the counter-ion system corresponds to a bilayer
Wigner crystal, as a consequence of Earnshaw theorem \cite{Earnshaw}. 
The lattice spacings of each layer are denoted $b_1$ and $b_2$;
they are the direct counterpart of the length scale $a$ introduced in 
section \ref{sec:oneplate}. 
The bilayer structure is, in general, complicated and depends on 
the distance $d$ \cite{Goldoni96,Messina03,Lobaskin07}.
For this region of $\zeta$ values, the WSC technique cannot be adapted 
directly from the one-plate geometry, except for small distances between 
the plates such that $d\ll b$, where $b=\min\{b_1,b_2\}$.
The point is that each particle experiences,
besides the direct linear one-body potential (\ref{eq:one-body}) induced by 
homogeneously charged plates, an additional perturbation due to the
repulsive interactions with other $q$-valent ions.
This additional potential is,  for $d\ll b$,
small compared to (\ref{eq:one-body}). This opens the way to a perturbative
treatment along similar lines as in section \ref{sec:oneplate}, in which
the leading one-body description is then fully equivalent to the one derived within
the VSC method.

First we shall address the symmetric $\zeta=1$ case which ground state was 
studied extensively in the past.
The symmetric configuration is of special importance in the VSC method: 
Although the leading SC result for the density profile and 
the pressure was derived for all values of the asymmetry parameter 
$-1\le \zeta \le 1$ \cite{Kanduc08}, the first SC correction 
(inconsistent with MC simulations) is available up to now only 
for $\zeta=1$ \cite{Moreira00,Netz01}. 
After solving the SC limit for the symmetric case, we shall pass 
to asymmetric, oppositely and likely charged, surfaces and solve
the problem in the leading SC order plus the first correction.

\subsection{Equivalently charged plates}
\begin{figure}[htb]
\begin{center}
\includegraphics[width=0.35\textwidth]{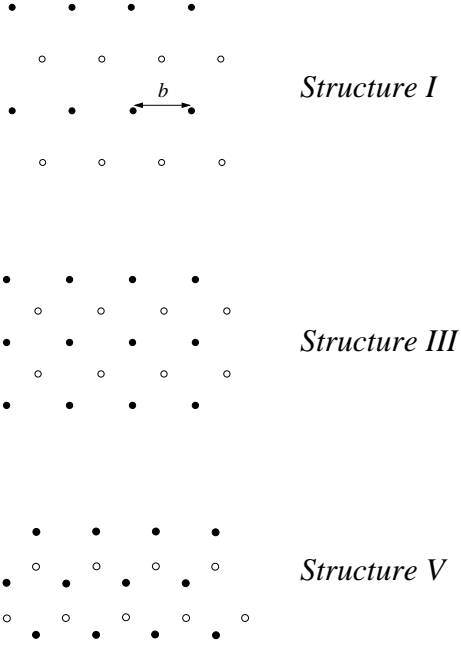}
\caption{Rigid ground-state structures I, III and V of counter-ions on 
two parallel charged plates; open and filled symbols correspond to particle
positions on the opposite surfaces.}
\label{fig:Structures}
\end{center}
\end{figure}

For $\sigma_1=\sigma_2=\sigma$, the electric field between the walls 
vanishes. 
At $T=0$, the classical system is defined furthermore by the dimensionless 
separation 
\begin{equation} \label{eq:eta}
\eta = d \sqrt{\frac{\sigma}{q}} = \frac{1}{\sqrt{2\pi}}
\frac{\widetilde{d}}{\sqrt{\Xi}} .
\end{equation} 
A complication comes from the fact that counter-ions form, on the opposite 
surfaces, a bilayer Wigner crystal, the structure of which depends on $\eta$ 
\cite{Goldoni96,Messina03,Lobaskin07}.
Two limiting cases are clear. 
At the smallest separation $\eta=0$, a single hexagonal Wigner crystal 
is formed. 
Due to global neutrality, its lattice spacing $b$ is given by
\begin{equation} \label{eq:spacing}
\frac{q}{2\sigma} = \frac{\sqrt{3}}{2} b^2 .
\end{equation}
The lattice spacing is simply related to that of the one plate problem 
by $b=a/\sqrt{2}$.
At large separations $\eta\to \infty$, each of the plates has its own
Wigner hexagonal structure and these structures are shifted with respect
to one another.
The transition between these limiting phases corresponds to
the following sequence of structures (in the order of increasing $\eta$ \cite{Goldoni96}):
a mono-layer hexagonal lattice (I, $0\le\eta\le \eta_0$), 
a staggered rectangular lattice (II, $\eta_0<\eta\le 0.26$),
a staggered square lattice (III, $0.26<\eta\le 0.62$), 
a staggered rhombic lattice (IV, $0.62<\eta\le 0.73$) and
a staggered hexagonal lattice (V, $0.73<\eta$).
The three ``rigid'' structures I, III and V, which do not change within 
their stability regions, are shown in Fig. \ref{fig:Structures}.
The primary cells of intermediate ``soft'' II and IV lattices are changing
with $\eta$ within their stability regions.
The existence of phase I in a small, but finite interval of $\eta$,
is a controversial issue \cite{Goldoni96,Messina03,Lobaskin07},
and therefore, so is the case of the precise value of the threshold $\eta_0$.
Whether $\eta_0$ is vanishing or is a very small number, remains an open problem.
Here, we perform expansions of thermodynamic quantities in powers
of $d/b\ll 1$ (or, equivalently, $\eta \propto\widetilde{d}/\sqrt{\Xi}\ll 1$
since the scale $\widetilde{d}$ is fixed while $\Xi$ becomes large).
We therefore need to know the ground state structure for $d/b\propto \eta=0$,
which is clearly structure I, irrespective of the ``$\eta_0$ controversy'', 
with a lattice spacing given by (\ref{eq:spacing}).
We shall thus document our WSC expansion on structure I.

Let ${\bf R}_j=(R_j^x,R_j^y)$ be the position vector of the particle localized 
on the shared hexagonal Wigner lattice of type I; $Z_j=0$ if the particle
$j=1,\ldots,N/2$ belongs to the plate $\Sigma_1$ (say filled symbols of 
Structure I in Fig. \ref{fig:Structures}) and $Z_j=d$ if the particle
$j=N/2+1,\ldots,N$ belongs to the plate $\Sigma_2$ (open symbols of 
Structure I in Fig. \ref{fig:Structures}).
Let us shift all particles from their lattice positions 
$\{ {\bf R}_j,Z_j=0\vee d \}$ to $\{ (x_j,y_j,z_j) \}$ and look for the 
corresponding energy change $\delta E$ from the ground state.
Since the potential induced by the surface charge on the walls is
constant between the walls and the linear in $z$ contribution of Wigner 
crystals is negligible if $d/b\ll 1$, the corresponding $\delta E^{(1)}=0$.
The $z$-coordinates of particles, constrained by the distance $d$ between 
the plates, are much smaller than the Wigner lattice spacing $b$, 
i.e. both $d^2$ and $(z_j-z_k)^2$ are much smaller than 
$\vert {\bf R}_j-{\bf R}_k\vert^2$ for $j\ne k$.
The harmonic in $z$ part of the energy change thus reads
\begin{eqnarray} \label{eq:deltainteraction}
\delta E^{(2)}_z & = & - \frac{(qe)^2}{4\varepsilon}
\sum_{j,k=1\atop (j\ne k)}^N \frac{(z_j- z_k)^2}{
\vert {\bf R}_j-{\bf R}_k \vert^3}  \nonumber \\ & & 
+ \frac{(qe)^2}{2\varepsilon} \sum_{j\in\Sigma_1} \sum_{k\in\Sigma_2} 
\frac{d^2}{ \vert {\bf R}_j-{\bf R}_k\vert^3} .
\end{eqnarray}
Note that the first (quadratic in $z$) term carries only the information 
about the single Wigner crystal of lattice spacing $b$.
The information on how the lattice sites are distributed between the two
plates within structure I is contained in the second constant
(from the point of view of thermal averages irrelevant) term which compensates 
the first one if the counter-ions are in their ground-state configuration. 
The harmonic terms in the $(x,y)$ plane prove immaterial for the sake of
our purposes.
The total energy change is given, as far as the $z$-dependent contribution 
is concerned, by $-\beta\delta E = S_z$ with
\begin{eqnarray}
S_z \sim \frac{\left( \sqrt{2}\alpha \right)^3}{4\sqrt{\Xi}}
\sum_{j,k=1\atop (j\ne k)}^N \frac{(\widetilde{z}_j-\widetilde{z}_k)^2}{(R_{jk}/b)^3}  . 
\end{eqnarray}
The only difference between this two-plate $S_z$ and the one-plate $S_z$
(\ref{eq:Sz}) consists in the factor $2^{3/2}$ due to the different lattice
spacing of the corresponding Wigner crystals, $b=a/\sqrt{2}$.

To derive the density profile, we use the cumulant technique with the
one-body Boltzmann factor equal to 1 (no external potential).
The leading SC behavior stems from
$Z_N^{(0)}[w] = \left[ \int_{\Lambda}d{\bf r}w({\bf r})\right]^N/N!$.
Since
\begin{equation}
\frac{\delta}{\delta w({\bf r})} \ln Z_N^{(0)}[w] \Big\vert_{w({\bf r})=1}
= \frac{N}{\vert\Sigma\vert d} = (2\pi\ell_{\rm B}\sigma^2) \frac{2}{\widetilde{d}}
\end{equation}
we have in the leading SC order the constant density
$\widetilde{\rho}_0(\widetilde{z})\sim 2/\widetilde{d}$.
This is the one-particle result in zero potential, respecting the
electroneutrality condition (\ref{eq:electro2}) with $\zeta=1$.
The same leading form was obtained by the VSC method \cite{Moreira00,Netz01}.
The physical meaning is simple: due to their strong mutual repulsion, 
the counter-ions form a strongly modulated structure along the plate
and consequently decouple in the transverse direction, where they only 
experience the electric field due to the two plates. In the symmetric case
$\zeta=1$, this field vanishes and the resulting ionic density 
is uniform along $z$: from electroneutrality, it reads 
$\widetilde \rho_0 = 2/\widetilde d$. The situation changes
in the asymmetric case, where one can anticipate $\widetilde \rho_0$,
again driven by the non vanishing but 
uniform bare plates field, to be exponential in $z$.

The first correction to the density profile stems from
\begin{equation}
\langle S_z \rangle_0 \sim \frac{\sqrt{2}\alpha^3}{\sqrt{\Xi}} N C_3
\left( [\widetilde{z}^2]_0 - [\widetilde{z}]_0^2 \right) ,
\end{equation}
where
\begin{equation}
[\widetilde{z}^p]_0 \equiv \frac{\int_{\Lambda}d{\bf r} w({\bf r}) \widetilde{z}^p}{
\int_{\Lambda}d{\bf r} w({\bf r})} , \quad p=1,2,\ldots. 
\end{equation}
Simple algebra yields
\begin{equation}
\frac{\delta}{\delta w({\bf r})} [\widetilde{z}^p]_0 \Big\vert_{w({\bf r})=1}
= \frac{1}{\vert\Sigma\vert d} \left( \widetilde{z}^p 
- \frac{\widetilde{d}^p}{p+1} \right) ,
\end{equation}
where we used that 
$[\widetilde{z}^p]_0\vert_{w({\bf r})=1} = \widetilde{d}^p/(p+1)$.
The density profile $\widetilde{\rho}(\widetilde{z})$ is thus obtained in the form
\begin{equation} \label{eq:profile}
\widetilde{\rho}(\widetilde{z}) = \frac{2}{\widetilde{d}} + \frac{1}{\theta}
\frac{2}{\widetilde{d}} \left[ \left( \widetilde{z}-\frac{\widetilde{d}}{2} \right)^2
- \frac{\widetilde{d}^2}{12} \right] + {\cal O}\left( \frac{1}{\Xi} \right) ,
\end{equation}
where
\begin{equation} \label{eq:theta2}
\theta(\zeta=1) = \frac{(4\pi)^{3/2}}{3^{3/4}} 
\frac{1}{C_3} \frac{1}{\sqrt{2}} \sqrt{\Xi}
= 1.252\ldots \sqrt{\Xi} .
\end{equation}
This density profile respects the electroneutrality condition
(\ref{eq:electro2}) with $\zeta=1$.
The functional form of (\ref{eq:profile}) coincides with that of Moreira 
and Netz \cite{Moreira00,Netz01}.
For (not yet asymptotic) $\Xi=100$, the previous VSC result $\theta=\Xi$ 
is far away from the MC estimate $\theta\simeq 11.2$ \cite{Moreira00}, 
while our formula (\ref{eq:theta2}) gives a reasonable value 
$\theta\simeq 12.5$.

In the evaluation of the $\theta$ factor in Eq. (\ref{eq:theta2}),
we use the exact result (\ref{eq.C3}) for the lattice sum $C_3$ 
of the mono-layer hexagonal structure I, which was the starting point 
of our expansion.
It is instructive to compare (\ref{eq:theta2}) with the corresponding
$\theta$ factors calculated for the structures III and V presented in
Fig. \ref{fig:Structures}.
Using a representation of the lattice sums in terms of quickly convergent
integrals over products of Jacobi theta functions, we find that 
$\theta=1.232\ldots \sqrt{\Xi}$ for the structure III and
$\theta=1.143\ldots \sqrt{\Xi}$ for the structure V.
These values show only a slight dependence of $\theta$ on the 
structure of the ground state.

\begin{figure}[htb]
\begin{center}
\includegraphics[width=0.45\textwidth,clip]{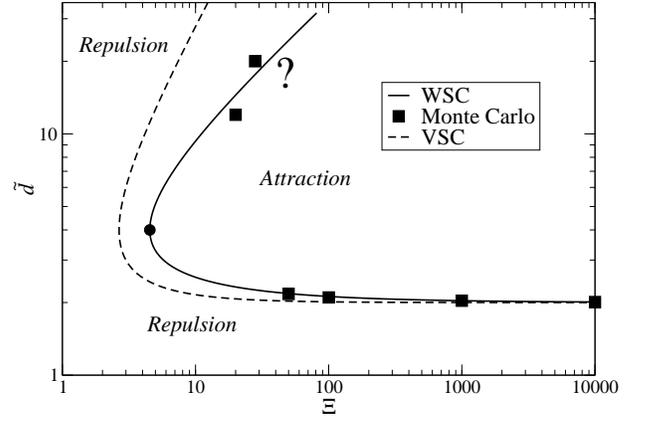}
\caption{Phase diagram following from the WSC equation of state (\ref{eq:eos}),
for symmetric like-charged plates ($\zeta=1$).
The solid curve, which shows the points where $P=0$, divides the
$(\Xi,\widetilde{d})$ plane onto its attractive $(P<0)$ and repulsive
$(P>0)$ parts.
The dashed line is the original VSC prediction \cite{Netz01}. 
The filled squares are the MC data from Ref. \cite{Moreira00} with $\Xi>20$.
The filled circle indicates the terminal point of the 
attraction/repulsion separatrix, obtained within WSC.
The question mark is a reminder that the upper branch of the isobaric curve 
$P=0$ is such that $\widetilde d \propto \sqrt{\Xi}$, whereas our results
are meaningful under the proviso that $\widetilde d \ll \sqrt{\Xi}$.}
\label{fig:phases}
\end{center}
\end{figure}

Applying the contact-value theorem (\ref{eq:contacttheorem2}) to the density 
profile (\ref{eq:profile}), the pressure $P$ between the plates is given by 
\begin{equation} \label{eq:eos}
\widetilde{P} = - 1 + \frac{2}{\widetilde{d}} + \frac{\widetilde{d}}{3\theta}
+ {\cal O}\left(\frac{1}{\Xi}\right) .
\end{equation}
A similar result was obtained within the approximate approach 
of Ref. \cite{Hatlo10}, with the underestimated ratio 
$\theta/\sqrt{\Xi}=3\sqrt{3}/2=0.866\ldots$.

Equation (\ref{eq:eos}) provides insight into the like charge attraction phenomenon.
The attractive ($P<0$) and repulsive ($P>0$) regimes are shown in 
Fig. \ref{fig:phases}.
Although our results hold for $\widetilde{d} \ll \sqrt{\Xi}$ and for large $\Xi$, 
the shape of the phase boundary where $P=0$ (solid curve) shows striking 
similarity with its counterpart obtained numerically \cite{Moreira00,Chen06}. 
For instance, the terminal point of the attraction region, shown by 
the filled circle in Fig. \ref{fig:phases}, is located at $\widetilde{d}=4$, 
a value close to that which can be extracted from \cite{Moreira00,Chen06}. 
However, for $\Xi<20$, our results depart from the MC data, and in particular,
WSC underestimates the value of $\Xi$ at the terminal point:
we find $\Xi_{term} \simeq 4.53$ (corresponding to a critical value 
$\theta_{term}$=8/3), whereas the numerical data reported in \cite{Moreira00}
yields  $\Xi_{term} \simeq 12$. The previous results apply to the
VSC approach as well, where the functional form of the equation of
state is the same as in WSC. Since we have $\theta=\Xi$ in VSC, we conclude
that $\Xi_{term} = 8/3 \simeq 2.66$ within VSC, which is indeed the
value that can be seen in Fig. \ref{fig:phases}. Clearly, accounting
correctly for the behaviour of the counter-ion mediated pressure 
for $\Xi\leq 20$ requires to go beyond the strong-coupling analysis.
In addition, one has to be cautious as far as the location of the upper branch of 
the attraction/repulsion boundary is concerned: It is such that $\widetilde d/\sqrt{\Xi}$ is of 
order unity and hence lies at the border of validity of our expansion. 

\begin{figure}[htb]
\begin{center}
\includegraphics[width=0.45\textwidth,clip]{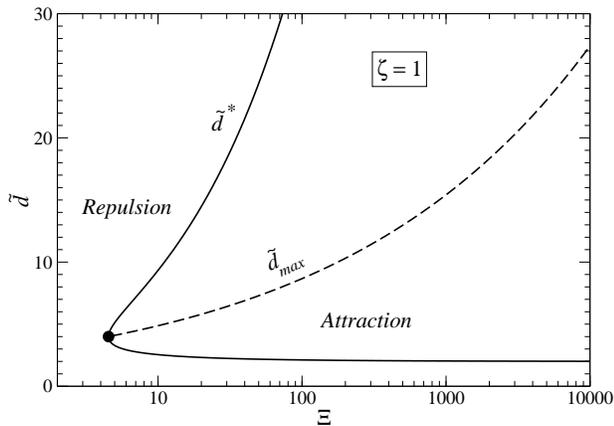}
\caption{The symmetric case $\zeta=1$: The maximum attraction distance 
$\widetilde{d}_{\text{max}}$ (dashed line) is defined by 
$\partial\widetilde{P}/\partial\widetilde{d}=0$. 
The solid curve $\widetilde{d}^*$ is the boundary between attractive 
and repulsive regimes.}
\label{fig:dmax}
\end{center}
\end{figure}

There is another feature of the equation of state under strong
coupling that can be captured by our analysis: The distance of maximal
attraction, where the pressure is most negative. 
We predict the maximum attraction, following from  $\partial\widetilde{P}/\partial\widetilde{d}=0$,
to be reached at
$\widetilde{d}_{\text{max}} = \sqrt{6 \theta} \propto \Xi^{1/4}$. 
Since $\widetilde{d}_{\text{max}}/\sqrt{\Xi} \propto \Xi^{-1/4} \to 0$
in the asymptotic limit $\Xi\to\infty$, we can consider the latter prediction, 
shown by the dashed line in Fig. \ref{fig:dmax}, as asymptotically exact. 
We note that it is fully corroborated by the scaling laws reported 
in \cite{Chen06}, while VSC yields the scaling behaviour 
$\widetilde{d}_{\text{max}}  \propto \Xi^{1/2}$.

\begin{figure}[htb]
\begin{center}
\includegraphics[width=0.45\textwidth,clip]{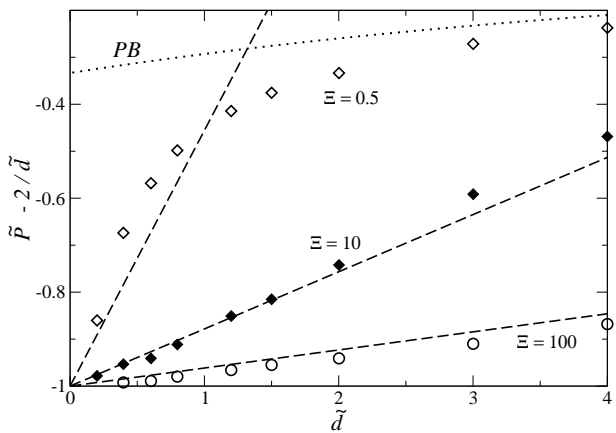}
\caption{The dependence of $\widetilde{P}-2/\widetilde{d}$ on the plate separation
$\widetilde{d}$ for three values of the coupling constant $\Xi=100$, $10$ 
and $0.5$. Here $\zeta=1$ (symmetric case).
The plots yielded by the WSC equation of state (\ref{eq:eos}) are
represented by dashed lines. 
Monte Carlo data \cite{Moreira00} are shown with symbols: open circles for $\Xi=100$, 
filled diamonds for $\Xi=10$ and open diamonds for $\Xi=0.5$. For completeness,
the Poisson-Boltzmann prediction is provided (dotted line in the upper part of the graph).}
\label{fig:Moreira16}
\end{center}
\end{figure}

We now analyze in more details the short distance behaviour of the
pressure.
The difference $\widetilde{P}-2/\widetilde{d}$, which is equal to $-1$ in 
the leading SC order and is linear in $\widetilde{d}$ as concerns 
the first correction, is plotted in Fig. \ref{fig:Moreira16} as 
a function of the (dimensionless) plate separation $\widetilde{d}$.
Three values of the coupling constant were considered: $\Xi=100$, $10$ and $0.5$.
The plots obtained from the equation of state (\ref{eq:eos}) are
shown by dashed lines and the MC data \cite{Moreira00}
are represented by symbols.
The accuracy of the WSC method is good, surprisingly also
for small values of $\Xi=10$ and $0.5$, where the approach is not supposed to hold. 
As concerns the (leading term plus the first correction) VSC equation of 
state \cite{Netz01}, corresponding to our Eq. (\ref{eq:eos}) with $\theta=\Xi$, 
the plots for $\Xi=10$ and $100$ are close to the $\widetilde{d}$ axis, 
and far from the Monte Carlo data; we consequently do not present them in the figure.
For $\Xi=0.5$, the VSC prediction is in good agreement with the 
MC simulations \cite{Moreira00}. It is interesting to note that in the distance range
$\widetilde d < 2$, the $\Xi=0.5$ data depart from the mean-field
(Poisson-Boltzmann) results \cite{Moreira00}, see Fig. \ref{fig:Moreira16}: there, the inter-plate
distance becomes comparable or smaller to $b$, which means that
the discrete nature of the particles can no longer be ignored;
At larger distances only does the continuum mean-field description
hold. For small inter-plate distances, we expect the single
particle picture to take over, no matter how small $\Xi$ is.
This explains that $\widetilde{P}-2/\widetilde{d}\to -1$,
but there is then no reason that WSC or VSC would provide the
relevant $\widetilde d$ correction at small $\Xi$.
The fact that WSC and VSC agree with each other here at $\Xi=0.5$ is a hint
that such a correspondence with MC is incidental
(and indeed, in this range of couplings, $\Xi$ and $\Xi^{1/2}$ are of the 
same order). It would be
interesting to have MC results at very small $\Xi$ values,
and to concomitantly develop a theory for the first pressure correction
to the leading term $2/\widetilde{d}-1$.

\subsection{Asymmetrically charged plates}
The sequence of ground states for asymmetric like-charged plates 
$(0<\zeta\le 1)$ may be even more complex than the one for the symmetric 
$\zeta=1$ case; in dependence on the distance $d$, the bilayer Wigner crystal 
can involve commensurate as well as incommensurate structures of counter-ions.
In addition, related work in spherical geometry  \cite{Messina09,Messina00} has shown that the ground state
in general breaks local neutrality (the two partners acquire
an electrical charge, necessarily opposite).
The possibility of, in principle, an infinite number of irregular structures 
might complicate numerical calculations;
we are not aware about a work dealing with this subject.

\begin{figure}[htb]
\begin{center}
\includegraphics[width=0.45\textwidth,clip]{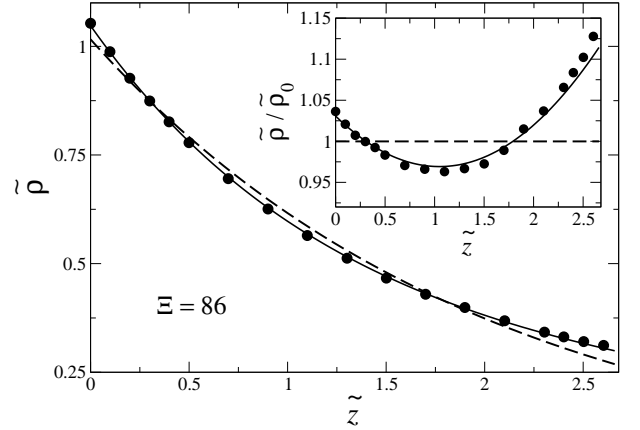}
\caption{The density profile $\widetilde{\rho}$ for $\zeta=0.5$, $\Xi=86$ and 
$\widetilde{d}=2.68$. The dashed curve corresponds to the leading SC profile
$\widetilde{\rho}_0$ (\ref{eq:leadingscprof}), the solid curve also involves
the first correction in (\ref{eq:totalprof}). 
MC data (filled circles) come from Ref. \cite{Kanduc08}. The inset shows the ratio 
$\widetilde\rho/\widetilde\rho_0$, for a finer test of the correction to leading
order $\widetilde\rho_0$. }
\label{fig:n_Kanduc_dz0.5}
\end{center}
\end{figure}

\begin{figure}[htb]
\begin{center}
\includegraphics[width=0.45\textwidth,clip]{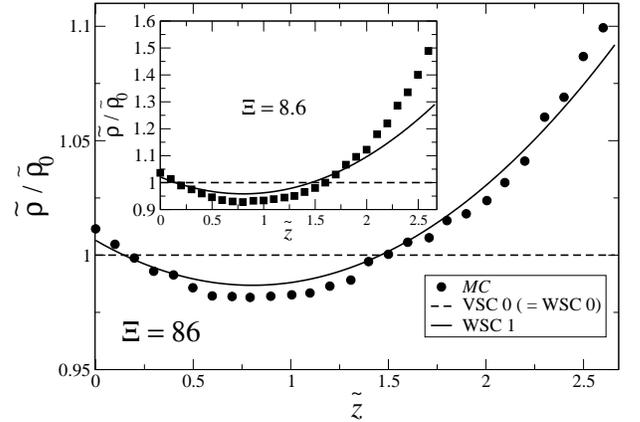}
\caption{Same as in the inset of Fig. \ref{fig:n_Kanduc_dz0.5},
for $\zeta=0$, and two different values of the coupling
parameter $\Xi$. 
The two plates are located at $\widetilde z=0$ and $\widetilde z = 2.68$.
Here, $\zeta=0$ means that the plate at $\widetilde z = 2.68$ is uncharged.
The symbols are for the Monte Carlo data of Ref. \cite{Kanduc08}.}
\label{fig:n_Kanduc_dz0}
\end{center}
\end{figure}

Fortunately, the same simplification as for the equivalently charged plates arises
at small separations between the plates $d/b\ll 1$, where the lateral 
lattice spacing $b$ of the single Wigner crystal is now given by 
the requirement of the global electroneutrality, as follows:
\begin{equation} \label{eq:spacingb}
\frac{q}{\sigma_1+\sigma_2} = \frac{\sqrt{3}}{2} b^2 .
\end{equation} 
Since the $z$-coordinates of particles between the plates are much smaller 
than $b$, we can use the harmonic $z$-expansion of the interaction
energy of type (\ref{eq:deltainteraction}), where only the (irrelevant) 
constant term reflects the formation of some nontrivial bilayer structure.
Our task is to derive the particle density profile for 
the energy change from the ground state of the form
\begin{equation}
- \beta \delta E = - \kappa \sum_{j=1}^N \widetilde{z}_j + S_z ,
\end{equation}
where $\kappa = 1-\zeta=1-\sigma_2/\sigma_1$ and 
\begin{eqnarray}
S_z & \sim & \frac{q^2 \ell_{\rm B}}{4} \sum_{j,k=1\atop (j\ne k)}^N
\frac{(z_j-z_k)^2}{\vert {\bf R}_j-{\bf R}_k\vert^3} \nonumber \\
& = & \frac{\left( \sqrt{1+\zeta}\alpha \right)^3}{4\sqrt{\Xi}}
\sum_{j,k=1\atop (j\ne k)}^N \frac{(\widetilde{z}_j-\widetilde{z}_k)^2}{(R_{jk}/b)^3} . 
\end{eqnarray}
We use the cumulant technique with the one-body Boltzmann factor
$\exp(-\kappa\widetilde{z})$.
The final result for the density profile reads
\begin{eqnarray} \label{eq:totalprof}
\widetilde{\rho}(\widetilde{z}) & = & (1-\zeta^2)
\frac{e^{-\kappa\widetilde{z}}}{1-e^{-\kappa\widetilde{d}}} \Bigg\{ 1 + 
\frac{\left( \sqrt{1+\zeta}\alpha \right)^3 C_3}{2\sqrt{\Xi}} \nonumber \\
& & \times \left[ \widetilde{z}^2 - t_2 - 2 t_1 (\widetilde{z}-t_1) \right] 
+ {\cal O}\left( \frac{1}{\Xi} \right) \Bigg\} ,
\end{eqnarray}
where
\begin{eqnarray} \label{eq:t1t2}
t_1(\kappa) & = & \frac{\int_0^{\widetilde{d}} d\widetilde{z} \widetilde{z} 
e^{-\kappa\widetilde{z}}}{\int_0^{\widetilde{d}} d\widetilde{z} e^{-\kappa\widetilde{z}}} 
= \frac{1}{\kappa} - \frac{\widetilde{d}}{e^{\kappa\widetilde{d}}-1} , \\
t_2(\kappa) & = & \frac{\int_0^{\widetilde{d}} d\widetilde{z} \widetilde{z}^2 
e^{-\kappa\widetilde{z}}}{\int_0^{\widetilde{d}} d\widetilde{z} e^{-\kappa\widetilde{z}}} \nonumber \\
& = & \frac{2}{\kappa^2} - \frac{1}{e^{\kappa\widetilde{d}}-1} 
\left( \frac{2\widetilde{d}}{\kappa} + \widetilde{d}^2 \right) .
\end{eqnarray}
For example, the density profile $\widetilde{\rho}$ for $\zeta=0.5$, $\Xi=86$ 
and $\widetilde{d}=2.68$ is depicted in Fig. \ref{fig:n_Kanduc_dz0.5}. 
The dashed curve corresponds to the leading SC profile
\begin{equation} \label{eq:leadingscprof}
\widetilde{\rho}_0(\widetilde{z}) = (1-\zeta^2)
\frac{e^{-\kappa\widetilde{z}}}{1-e^{-\kappa\widetilde{d}}} ,
\end{equation}
which is the same in both VSC and WSC theories.
For the parameters of Fig. \ref{fig:n_Kanduc_dz0.5},
the leading order profile reads
\begin{equation}
\widetilde{\rho}_0(\widetilde{z}) = \frac{3}{4} \frac{e^{-\widetilde{z}/2}}{1-e^{-1.34}} .
\end{equation}
The WSC profile (\ref{eq:totalprof}), involving also the first SC correction, 
is represented by the solid curve.
The filled circles are the MC data of Ref. \cite{Kanduc08}.
The ratio $\widetilde{\rho}/\widetilde{\rho}_0$, which is trivially equal to 1
in the leading SC order, is presented in the inset of the figure;
we see that the first correction improves substantially the agreement with MC data.
A similar conclusion is reached in the case where one plate is 
uncharged ($\zeta=0$), see Fig. \ref{fig:n_Kanduc_dz0}: for the highest
coupling investigated numerically in Ref. \cite{Kanduc08} ($\Xi=86$),
the agreement between the WSC approach and Monte Carlo data
for the density profile is excellent, and subtle deviations
from the leading order term $\rho_0$ are fully captured. 
It can be seen in the inset of Fig. \ref{fig:n_Kanduc_dz0}
that the agreement is no longer quantitative when the coupling
parameter is decreased by a factor of 10. As may have been anticipated,
the density profile close to the highly charged plate located
at $\widetilde z = 0$ is well accounted for by our treatment,
while the agreement with MC deteriorates when approaching the
uncharged plate located at $\widetilde z = \widetilde d$. We 
may anticipate that the WSC approach would fare better against Monte Carlo at smaller 
inter-plate separations.

Either of the contact-value relations (\ref{eq:contacttheorem2}) implies
the same pressure:
\begin{equation} \label{eq:p}
\widetilde{P} = \widetilde{P}_0 + \frac{1}{\sqrt{\Xi}} \widetilde{P}_1 
+ {\cal O}\left( \frac{1}{\Xi} \right) ,
\end{equation}
where
\begin{equation} \label{eq:p0}
\widetilde{P}_0 = - \frac{1}{2} (1+\zeta^2) + 
\frac{1}{2} (1-\zeta^2) \coth\left( \frac{1-\zeta}{2}\widetilde{d} \right)
\end{equation}
is the leading SC contribution, already obtained within the VSC
method in \cite{Kanduc08}, and
\begin{eqnarray}
\widetilde{P}_1 & = & \frac{3^{3/4} (1+\zeta)^{5/2} C_3}{4 (4\pi)^{3/2}} 
\frac{\widetilde{d}}{\sinh^2\left(\frac{1-\zeta}{2}\widetilde{d} \right)} \nonumber \\
& & \times \left[ \left( \frac{1-\zeta}{2}\widetilde{d} \right) 
\coth\left( \frac{1-\zeta}{2}\widetilde{d} \right) - 1 \right] \label{eq:p1}
\end{eqnarray}
is the coefficient of the first $1/\sqrt{\Xi}$ correction.

\begin{figure}[htb]
\begin{center}
\includegraphics[width=0.45\textwidth,clip]{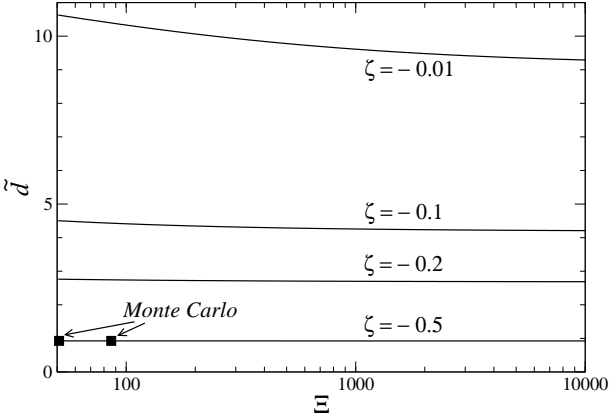}
\caption{Oppositely charged plates: The phase boundary where $\widetilde{P}=0$,
which discriminates the attractive regime (at large distances) from the 
repulsive one (at small distances). The MC data for $\zeta=-0.5$ (filled squares) come from Ref. \cite{Kanduc08}.}
\label{fig:dzmoins}
\end{center}
\end{figure}

\begin{figure}[htb]
\begin{center}
\includegraphics[width=0.45\textwidth,clip]{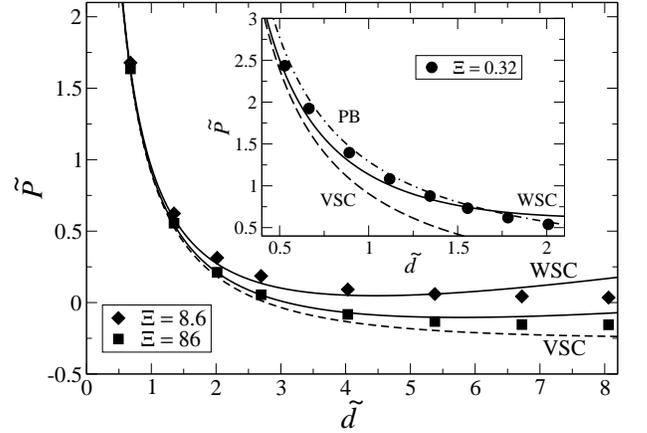}
\caption{Rescaled pressure versus the plate distance for likely charged 
plates with the asymmetry parameter $\zeta=0.5$:
The dashed curve corresponds to the leading term of the VSC theory, which
is equivalent to the WSC one (\ref{eq:p0}).
The small-$\widetilde{d}$ expansion of the WSC pressure (\ref{eq:stateasym})
is represented by solid curves.
Filled symbols represent the MC data \cite{Kanduc08} for the couplings
$\Xi=86$ (squares), $\Xi=8.6$ (diamonds) and $\Xi=0.32$ 
(circles in the inset). In the inset, which is a zoom on the
small distance region, the mean-field Poisson-Boltzmann
(PB) prediction is also displayed.}
\label{fig:dzpluss}
\end{center}
\end{figure}

While the first correction to the pressure $\widetilde{P}_1$ vanishes in 
both limits $\widetilde{d}\to 0$ and $\widetilde{d}\to\infty$, 
$\widetilde{P}_0$ is in general nonzero and therefore dominates in these 
asymptotic regions.
Let us first consider the large-$\widetilde{d}$ limit:
\begin{equation} \label{eq:infty}
\lim_{\widetilde{d}\to\infty} \widetilde{P} = \lim_{\widetilde{d}\to\infty} \widetilde{P}_0 
= - \zeta^2 .
\end{equation} 
Such a result is correct for oppositely charged plates $-1<\zeta\le 0$.
In that case indeed, for sufficiently distant plates, all counter-ions stay
in the neighborhood of plate 1 and compensate partially its surface
charge, that is reduced from the bare value $\sigma_1e$ to $\vert\sigma_2\vert e$.
We are left with a capacitor of opposite surface charges
$\pm \sigma_2 e$ whose dimensionless pressure is attractive and
just equal to $-\zeta^2$. In other words, again for large distances,
the negative counter-ions are expelled from the vicinity of the
negatively charged plate 2, with a resulting vanishing charge density
$\widetilde\rho(\widetilde d)$. From the contact theorem, this implies
that the pressure reads $\widetilde P = -\zeta^2$. Hence, the
leading SC order (common to VSC and WSC), {\it a priori} valid
at short distances, yields the correct result at large distances also.
This points to the adequacy of the WSC result (\ref{eq:p})-(\ref{eq:p1}) 
in the whole range of $\widetilde{d}$ values for oppositely charged plates, 
which is consistent with our previous analysis about the simple nature
of the ground state (independent on the inter-plate distance, at variance
with the $\zeta>0$ case). In addition, we emphasize that
the effect of the first correction coefficient (\ref{eq:p1}) is very weak.
This fact is documented in Fig. \ref{fig:dzmoins}: Each solid curve with 
a fixed asymmetry parameter $\zeta<0$ represents a phase boundary between
the anomalous repulsion of oppositely charged plates at small distances
and their ``natural'' attraction at large distances.
At $\Xi\to\infty$, using the condition $\widetilde{P}_0=0$ in (\ref{eq:p0})
implies the phase boundary at \cite{Kanduc08}
\begin{equation} \label{eq:dstar}
\widetilde{d}^* = - 2 \frac{\ln\vert\zeta\vert}{1-\zeta} , \qquad 
\mbox{$\Xi\to\infty$ \quad ($-1<\zeta<1$).}
\end{equation}
Considering also the first correction (\ref{eq:p1}) in (\ref{eq:p})
we see in Fig. \ref{fig:dzmoins} that the phase boundary $\widetilde{P}=0$ 
is almost independent of $\Xi$, except for very small negative values of 
$\zeta$. Consequently, the first correction to the leading SC behaviour
is generically negligible for oppositely charged plates.

\begin{figure}[htb]
\begin{center}
\includegraphics[width=0.45\textwidth,clip]{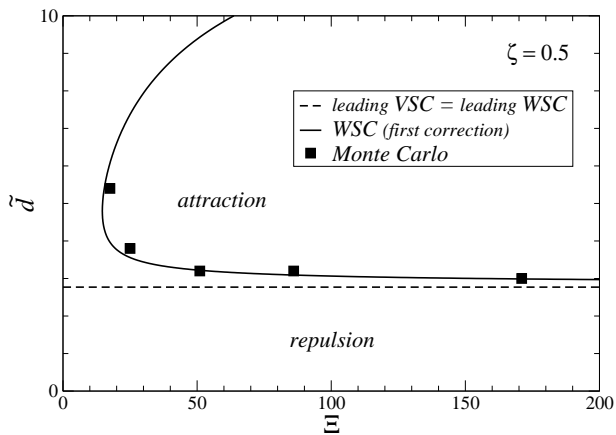}
\caption{Phase diagram for like-charged plates with asymmetry
parameter $\zeta=0.5$.
The phase boundary given by the leading VSC and WSC order \cite{Kanduc08} 
is represented by the dashed line.
The phase boundary following from our WSC result (\ref{eq:stateasym}) 
and (\ref{eq:theta3}) is represented by the solid curve;
for comparison, the filled squares are MC data from Ref. \cite{Kanduc08}.}
\label{fig:dtildedzplusKanduc}
\end{center}
\end{figure}

\begin{figure}[htb]
\begin{center}
\includegraphics[width=0.45\textwidth,clip]{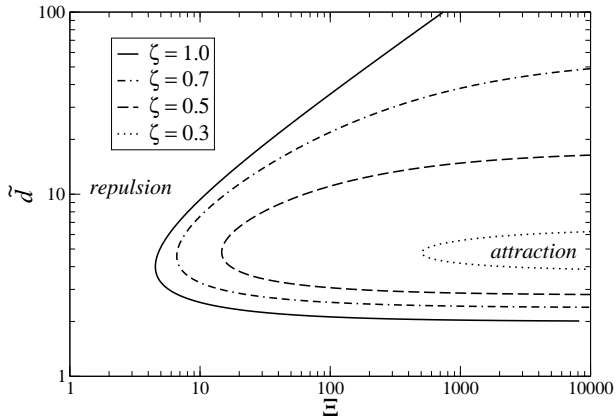}
\caption{The WSC phase boundaries for like-charged plates,
in the $(\Xi,\widetilde{d})$ plane and for various values of the asymmetry 
parameter $\zeta$.}
\label{fig:dzplus}
\end{center}
\end{figure}

On the other hand, the asymptotic result (\ref{eq:infty}) 
is apparently physically irrelevant for like-charged plates ($0<\zeta\le 1$). 
For sufficiently large distances $d$, the counter-ions stay in the neighborhood of 
both plates 1 and 2 and {\it a priori} neutralize their surface charges, so that 
the asymptotic pressure should vanish.
Therefore, for $\zeta>0$, we cannot expect the same bonus as for 
$\zeta<0$, and our WSC results (\ref{eq:p})-(\ref{eq:p1}) hold provided that 
$\widetilde{d}\ll\sqrt{\Xi}$ as was already the case for $\zeta=1$.
In addition, the small-$\widetilde{d}$ expansion of the pressure reads
\begin{eqnarray} \label{eq:stateasym}
\widetilde{P} & = & - \frac{1+\zeta^2}{2} + \frac{1+\zeta}{\widetilde{d}}
+ \left[ \frac{(1-\zeta)^2(1+\zeta)}{12} 
\right. \nonumber \\ & & \left. + \frac{1}{3\theta(\zeta)}
+ {\cal O}\left( \frac{1}{\Xi}\right) \right] \widetilde{d}
+ {\cal O}(\widetilde{d}^2) ,
\end{eqnarray}
where
\begin{equation} \label{eq:theta3}
\theta(\zeta) = \frac{(4\pi)^{3/2}}{3^{3/4}} 
\frac{1}{C_3} \frac{4}{(1+\zeta)^{5/2}} \sqrt{\Xi}.
\end{equation} 
As it should, this
is the generalization of the special $\zeta=1$ result (\ref{eq:theta2})
to all positive asymmetries. 

The plot of the rescaled pressure versus the plate distance for likely charged 
plates with the asymmetry parameter $\zeta=0.5$ is presented in 
Fig. \ref{fig:dzpluss}. 
The dashed curve corresponds to the leading term of the VSC theory, which
is equivalent to the leading WSC one (\ref{eq:p0}).
The small-$\widetilde{d}$ expansion of the WSC pressure (\ref{eq:stateasym})
is represented by solid curves.
The comparison with filled symbols of the MC data \cite{Kanduc08} 
shows a good agreement for the coupling constants $\Xi=86$ (squares), 
$\Xi=8.6$ (diamonds) and even for relatively small $\Xi=0.32$ 
(circles in the inset).
The agreement goes somewhat beyond the expected distance range of the validity
of the expansion (\ref{eq:stateasym}), but is restricted to the 
small $\widetilde d$ range.

The phase diagram for $\zeta=0.5$ is pictured in 
Fig. \ref{fig:dtildedzplusKanduc}.
The phase boundary given by the leading $\Xi\to\infty$ order of 
the VSC method \cite{Kanduc08} is represented by the dashed line.
As repeatedly emphasized above, it corresponds to the leading WSC order
as well.
The phase boundary following from our leading plus first correction
WSC result (\ref{eq:stateasym}) and (\ref{eq:theta3}) is represented by 
the solid curve; the agreement with MC data of Ref. \cite{Kanduc08} 
(filled squares) is very good.
The phase boundaries for like-charged plates with various values of 
the asymmetry parameter $\zeta$, following from our WSC result 
(\ref{eq:stateasym}) and (\ref{eq:theta3}), are drawn 
in the $(\Xi,\widetilde{d})$ plane in Fig. \ref{fig:dzplus}. 
It is seen that by decreasing $\zeta$ the anomalous attraction region
becomes smaller.

\begin{figure}[htb]
\begin{center}
\includegraphics[width=0.45\textwidth,clip]{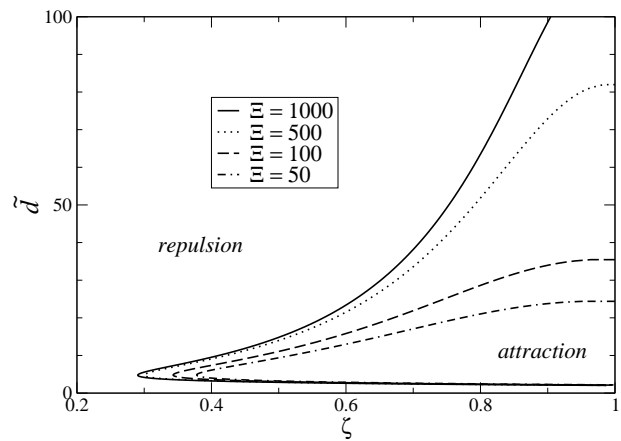}
\caption{The WSC phase boundaries for like-charged plates,
in the $(\zeta,\widetilde{d})$ plane and for various values of the coupling 
constant $\Xi$.}
\label{fig:dtildedzplus}
\end{center}
\end{figure}

\begin{figure}[htb]
\begin{center}
\includegraphics[width=0.45\textwidth,clip]{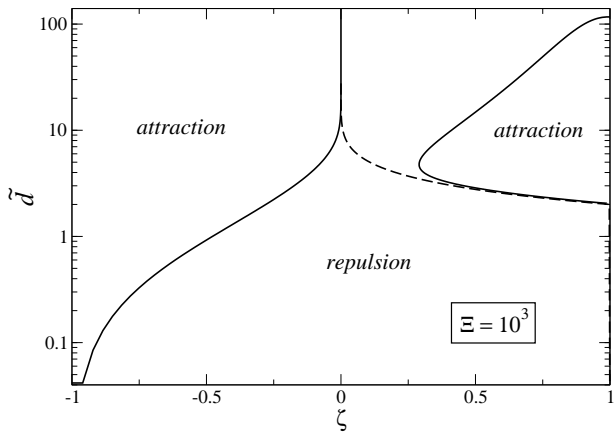}
\caption{The WSC phase diagram (solid curves) in the whole range of 
the asymmetry parameter $\zeta$, for the coupling constant $\Xi=10^3$.
For comparison, the phase diagram in the leading SC order (\ref{eq:dstar}) 
is represented by dashed curves; for oppositely charged plates $-1<\zeta\le 0$, 
the difference between the solid and dashed curves is invisible,
due to the already pointed out smallness of the first correction for $\zeta<0$.}
\label{fig:dtildedz}
\end{center}
\end{figure}

The WSC phase boundaries for like-charged plates, in the $(\zeta,\widetilde{d})$ 
plane and for various values of the coupling constant $\Xi$, are drawn 
in Fig. \ref{fig:dtildedzplus}.
For small values of the asymmetry parameter $\zeta$, e.g. below 
$\zeta\sim 0.29$ for $\Xi=10^3$, we see that the attractive ``pocket''
disappears. This phenomenon is entirely driven by the first correction,
as in revealed by Fig. \ref{fig:dtildedz}, which further shows the
phase diagram in the whole range of the asymmetry parameter $\zeta$, 
for the coupling constant $\Xi=10^3$.
For comparison, the phase boundaries between the repulsion and attractive
regions in the leading SC order, given by (\ref{eq:dstar}), are pictured by
dashed curves. With the corresponding leading contribution to the
pressure, the attractive region always exists.

\section{Conclusion}
\label{sec:concl}
In this paper, we have established the mathematical grounds for 
the Wigner Strong Coupling (WSC) theory which describes the strong-coupling regime 
of counter-ions at charged interfaces, 
starting from the Wigner structure formed at zero temperature.
The results for both likely and oppositely charged plates are in
excellent agreement with Monte Carlo data, which represents an improvement over the 
previously proposed Virial SC approach. By construction, our expansion should
be more reliable the larger the coupling parameter $\Xi$,
but we found that it remains trustworthy
for intermediate values of
the coupling constant (say $\Xi=100$), and in some cases down to $\Xi=10$
or 20. 

The geometries studied are those of one or two {\em planar} interfaces.
An important remark is that the leading results in the SC expansion follow from a single
counter-ion picture because the dominant (linear) electric potential stems
from the plate only; the contribution due to the interaction with
other counter-ions on the same plate is harmonic and therefore sub-dominant.
As a consequence, the leading terms of the VSC and WSC theories coincide.
This fact has been outlined on several occasions, but can nevertheless
not be considered as a general statement. Indeed, 
the situation changes for a {\em curved} (say, cylindrical or spherical)
wall surfaces since then the interactions of an ion with other counter-ions 
contribute to the dominant field, no matter how close to the interface 
this ion can be.
This is why the leading ion profile around a charged cylinder or sphere
will in general differ from that obtained within the original VSC approach \cite{Naji05}.
Inclusion of curvature effects in the WSC treatment 
is a task for the future.
In the present work, we have also assumed that the charges on the plates
are uniformly smeared, which opens the way to the powerful use of the
contact theorem to obtain the pressure. As a consequence, the 
interesting case of discrete fixed charges
on the plates \cite{MoreiraN02,Henle04,Trav06,Trav09}, 
is beyond the scope of the present
analysis.

A generalization of the formalism to quantum statistical systems of
counter-ions is straightforward: Vibrations of counter-ions around
their Wigner-lattice positions possess energy spectrum of quantized harmonic
oscillators.
Another perspective is to formulate a strong-coupling theory valid for an arbitrary 
distance between the plates. Indeed, both the original Virial SC and the present
Wigner SC theories are so far limited, in the two plate case, to the regime
$\widetilde d \ll \Xi^{1/2}$, which means that the inter plate distance 
should be smaller than the lattice spacing $a$ in the underlying Wigner crystal
(up to an irrelevant prefactor, the quantities $a$ and $b$ introduced in this
article refer to the same length).
It is important to emphasize here that the limitation $\widetilde d \ll \Xi^{1/2}$
is not intrinsic to the strong coupling limit, but is a technical requirement
that should be enforced to allow for the validity of the single particle
picture, and subsequent higher order corrections
as worked out here. Performing the SC expansion for distances $\widetilde d \gg \Xi^{1/2}$
requires to bypass the single particle picture, which is a challenging goal.
Finally, in view of possible applications to real colloidal systems,
it seems important to account for the low dielectric constant of 
colloidal particles, taking due account of image charge effects
\cite{Kanduc07,Levin11}. Work along these lines is in progress.

\begin{acknowledgments}
We would like to thank C. Texier for useful discussions.
L. \v{S}. is grateful to LPTMS for hospitality. 
The support received from the grants VEGA No. 2/0113/2009 and CE-SAS QUTE
is acknowledged. 
\end{acknowledgments}

\end{document}